\documentclass[aps,superscriptaddress,showpacs,floatfix,preprint]{revtex4}
\usepackage{times}
\usepackage{xspace}
\usepackage{graphicx,morefloats}
\usepackage{color}
\usepackage{amsmath, amsthm, amssymb}
\usepackage{multirow}
\def\degree{\hbox{$^\circ$}}

\begin{document}

\title{The Contribution of Internal and Model Variabilities to the Uncertainty in CMIP5 Decadal Climate Predictions}

\author{Ehud Strobach}
\author{Golan Bel}
\affiliation{Department of Solar Energy and Environmental Physics, Blaustein Institutes for Desert Research, Ben-Gurion University of the Negev, Sede Boqer Campus 84990, Israel}
\begin{abstract}
  Decadal climate predictions, which are initialized with observed conditions, are characterized by two main sources of uncertainties--internal and model variabilities.
Using an ensemble of climate model simulations from the CMIP5 decadal experiments, we quantified the total uncertainty associated with these predictions and the relative importance of each source.
Annual and monthly averages of the surface temperature and wind components were considered.
We show that different definitions of the anomaly results in different conclusions regarding the variance of the ensemble members.
However, some features of the uncertainty are common to all the measures we considered.
We found that over decadal time scales, there is no considerable increase in the uncertainty with time.
The model variability is more sensitive to the annual cycle than the internal variability. This, in turn, results in a maximal uncertainty during the winter in the northern hemisphere.
The uncertainty of the surface temperature prediction is dominated by the model variability, whereas the uncertainty of the wind components is determined by both sources.
Analysis of the spatial distribution of the uncertainty reveals that the surface temperature has higher variability over land and in high latitudes, whereas the surface zonal wind has higher variability over the ocean.
The relative importance of the internal and model variabilities depends on the averaging period, the definition of the anomaly, and the location.
These findings suggest that several methods should be combined in order to assess future climate prediction uncertainties and that weighting schemes of the ensemble members may reduce the uncertainties.
\end{abstract}
\maketitle

\section{Introduction}
Decadal climate predictions usually refer to predictions on time scales of one to three decades \cite{Smith10082007,Keenlyside2008,GRL:GRL29766,GRL:GRL51527,TELLUSA22830,Pohlmann2009,Kritman2013,meehl_decadal_2014}.
The meaningful products of these climate simulations are the averages of the climate variables over periods in the range of months to several years.
Accurate predictions of these temporal averages for several decades are of great interest to decision makers, agricultural producers and other stakeholders.
However, useful predictions must be accompanied by a measure of the uncertainty, i. e., the range of likely values of the predicted variables.
A common approach for estimating these uncertainties is by means of an ensemble of climate models \cite{Giorgi-grl-2000,Giorgi-cd-2000,raisanen_co2-induced_2001,Pan-jgr-2001,Giorgi-jc-2002,Webster-cc-2003,Jackson-jc-2004,Murphy2004,Cox13072007,Knutti2008,Jackson-jc-2008,deelia-cd-2008,Cretat-cd-2012,Knutti-cc-2012,Parker-wircc-2013,Yang-jgr-2013,Monier-gmd-2013,Solman-cd-2013,Blazquez-cd-2013,Zhao-acp-2013,Friedlingstein-jc-2014,Miao-ERL-2014,deser_projecting_2014,hawkins_potential_2009,hawkins_potential_2011,yip_simple_2011}.

On decadal time scales, uncertainties can be attributed to two main sources--internal and model variabilities \cite{hawkins_potential_2009}.
Internal variability is the spread of the climate predictions of the same climate model initialized with different, equally realistic, initial conditions (different realizations).
Model variability is the spread of the climate predictions of different models (we also include in this definition the sensitivity of each model to the various parameters).
A third source of uncertainty, which is more relevant in longer time scale predictions, is the scenario uncertainties.
These uncertainties reflect the variance of each model's climate predictions due to different projected atmospheric composition changes in the future.
Whereas internal variability is considered as an inherent noise of the climate models, which can hardly be reduced by ensemble methods, the model variability may be reduced \cite{Knutti-cc-2012} by weighting the ensemble models (e.g., based on their past performance) \cite{acpd-15-7707-2015}.
Therefore, quantification of the relative importance of each of these sources may be used to estimate the minimal uncertainties that can be achieved.
A good forecast is one for which the Root Mean Square Error (RMSE) of the prediction is equal to the uncertainty associated with the prediction \cite{palmer_2006} .
A forecast that has lower uncertainty than the RMSE is considered as overconfident.

Decomposing the uncertainty into its components should be done by using a reliable climate model ensemble composed of different realizations, models and scenarios; however, these are not always available. Different methods have been used to overcome the limited amount of data. The role of internal variability in the uncertainties of the CMIP2's (Coupled Model Intercomparison Project Phase 2) long-term climate projections was investigated \cite{raisanen_co2-induced_2001}. However, that data includes only one realization for each of the climate models. Therefore, the internal variability was estimated from the variance of segments of equal duration.
An analysis of global scale data \cite{Cox13072007} was extended to regional scales for surface temperature and precipitation using an ensemble taken from the CMIP3 experiment \cite{hawkins_potential_2009,hawkins_potential_2011}. This data also includes one realization for each model. To estimate the internal variability, they fitted the predictions of each of the models to a fourth degree polynomial, and the internal variability was defined as the variance of the differences between the polynomial fit and the model predictions. They showed that for the next few decades, the main sources of uncertainty are the model and the internal variabilities.
The uncertainties in climate predictions from the CMIP3 were decomposed in \cite{yip_simple_2011}.
In this analysis a full ensemble was used, including different models, different realizations of each model, and different scenarios, to decompose the prediction uncertainties into four components: internal, model, scenario and model-scenario interaction uncertainty.
The fourth component, which did not exist in the previous works mentioned, arose because of their definition of model and scenario uncertainties, which resulted in an interaction term.

The works discussed above (and others) describe the uncertainties of long-term climate projections. The common practice in these simulations is to initialize the climate models with a quasi-equilibrium steady state under a preindustrial atmospheric composition and to let them run into the future (usually until 2100) with observed past atmospheric composition changes and different scenarios for the future atmospheric composition changes. In these experiments, the outputs of interest are the 10-year (or longer) averages of the climate variables. The main interest is in estimating the response of the climate system to different atmospheric composition change scenarios. However, annual (or shorter period) averages of the climate variables are not expected to be synchronized with observations since they are initialized long before the present with somewhat arbitrary conditions \cite{meehl_decadal_2009}.
Therefore, a high resolution time series of the uncertainty derived from these projections is also not expected to be synchronized with the projections.

Here, we studied the uncertainties associated with near-term climate predictions, namely, decadal climate predictions from the CMIP5 \cite{taylor_overview_2012}.
The ensemble considered is composed of different models and realizations. This enables us to assess the relative importance of the internal and model variabilities.
The variables investigated here are the surface temperature (T) and the surface zonal wind ($U$). The surface meridional wind ($V$) was found to have similar characteristics to those of $U$, and therefore, the results of its analysis are provided in the Supplementary Materials.
The period of interest, 2006-2036, was chosen simply because it is the last decadal experiment of the CMIP5 that provides predictions for 30 years.
A list of the climate models (and the number of realizations of each model) included in the ensemble we analyzed is presented in Table \ref{CMIP5_models}.

\begin{table}
\caption{The CMIP5 decadal experiment models and the number of realizations for each of the models.}
\centering
\begin{tabular}{ccccc}
\hline\noalign{\smallskip}
Model & T & U/V& Resolution\\
\noalign{\smallskip}\hline\noalign{\smallskip}
bcc-csm1-1&4&4&64X128\\
CanCM4&20&20&64X128\\
CMCC-CM&3&3&240X480\\
FGOALS-g2&3&-&60X128\\
FGOALS-s2&3&3&108X128\\
HadCM3&10&10&73X96\\
IPSL-CM5A-LR&6&6&96X96\\
MIROC4h&6&6&320X640\\
MIROC5&6&6&128X256\\
MPI-ESM-LR&3&3&96X192\\
MRI-CGCM3&3&3&160X320\\
\noalign{\smallskip}\hline
\end{tabular}
\label{CMIP5_models}
\end{table}

The decadal climate simulations, unlike the CMIP5 long-term experiments, are initialized with observed ocean and sea ice conditions and, therefore, are expected to be synchronized with the climate system. The time series of the uncertainty is also expected to be synchronized with the predictions.
Decadal climate predictions were shown to produce a limited skill globally and a better skill in specific regions \cite{GRL:GRL29181,Doblas-Reyes2013,acpd-15-7707-2015}.
 
\section{Decomposing Uncertainties to Components}

The CMIP5 decadal predictions include different models and several realizations of each model. The effect of different scenarios is not significant over a decadal time scale, and therefore, different scenarios are not considered.
The method we use to decompose the sources of uncertainties is similar to those presented in \cite{raisanen_co2-induced_2001,yip_simple_2011}.
We decompose the uncertainty into its two main sources--internal and model variabilities.
Unlike previous studies \cite{yip_simple_2011}, we define the model and internal variabilities to be independent, and therefore, we do not have a model-internal ``interaction'' term.

For each time step and grid cell, we define the total variability of the ensemble predictions as the weighted ensemble variance, that is:

\begin{multline}
\label{eq:var}
\sigma_T^2 \equiv  \sum_{m=1}^N\sum_{r_m=1}^{N_m} w_{m,r_m}   \cdot (x_{m,r_m}- x_{\cdot,\cdot})^2
\\
= \sum_{m=1}^N\sum_{r_m=1}^{N_m} w_{m,r_m} \cdot  (x_{m,r_m} - x_{m,\cdot} + x_{m,\cdot} - x_{\cdot,\cdot})^2  
\\
=  \sum_{m=1}^N\sum_{r_m=1}^{N_m} w_{m,r_m} \cdot (x_{m,r_m} - x_{m,\cdot} )^2
\\
+ \sum_{m=1}^N\sum_{r_m=1}^{N_m} w_{m,r_m} \cdot ( x_{m,\cdot} - x_{\cdot,\cdot} )^2
\\
+2 \cdot   \sum_{m=1}^N \left( ( x_{m,\cdot}- x_{\cdot,\cdot}) \cdot  \sum_{r_m=1}^{N_m} w_{m,r_m}   \cdot  \left(x_{m,r_m} - x_{m,\cdot}\right)\right) 
\end{multline}
where $x_{m,r_m}$ is the predictions of the climate variable $x$ by the realization $r_m$ of model $m$, $w_{m,r_m}$ is its corresponding weight, $x_{m,\cdot}$ is the average of the realizations of model $m$, $x_{\cdot,\cdot}$ is the ensemble average (average of all the realizations and models), $N$ is the number of models and $N_m$ is the number of realizations of model $m$.
The last line in equation \ref{eq:var} is zero by definition, and therefore, equation \ref{eq:var} reads:

\begin{multline}
\label{eq:varc}
\sigma_T^2 \equiv \underbrace{\sum_{m=1}^N\sum_{r_m=1}^{N_m} w_{m,r_m} \cdot (x_{m,r_m} - x_{m,\cdot})^2}_{\sigma_I^2} \\
 +\underbrace{ \sum_{m=1}^N\sum_{r_m=1}^{N_m}  w_{m,r_m} \cdot (  x_{m,\cdot} - x_{\cdot,\cdot})^2 }_{\sigma_M^2}.
\end{multline}
This defines the two uncoupled contributions to the total variability, internal variability, $\sigma_I^2$, and model variability, $\sigma_M^2$.
For $N$ models and $N_m$ realizations of model $m$, we also define $w_{m,r} \equiv \frac{1}{N \cdot N_m}$ to avoid bias toward models with higher numbers of realizations.
Equation \ref{eq:varc} can then be slightly simplified:

\begin{multline}
\label{eq:varf}
\sigma_T^2 \equiv \underbrace{ \frac{1}{N}\sum_{m=1}^N\sum_{r_m=1}^{N_m} \frac{1}{N_m}\cdot (x_{m,r} - x_{m,\cdot})^2}_{\sigma_I^2}\\+\underbrace{ \frac{1}{N}\sum_m  \cdot ( x_{m,\cdot} - x_{\cdot,\cdot})^2 }_{\sigma_M^2}.
\\
\end{multline}

\section{Definition of the Anomaly}

It is known that climate models suffer from systematic bias, and a common practice is to correct their bias using different methods \cite{Xu2012,Bruyere_2014,meehl_decadal_2014}.
A simple, and the most frequently used, method is to subtract a constant factor (a different factor for each model), calculated from a reference period, from the predictions and analyze the spread of the anomalies.
In the long-term climate projections, it is the historical part of the experiment that is commonly used as a reference period.
The decadal experiments, on the other hand, do not have such a trivial historical reference period, and the definition of the anomalies is not trivial.
Moreover, it is also known that climate models initialized with observed conditions tend to drift to their preferred physical state during the first years after the initialization \cite{meehl_decadal_2009,meehl_decadal_2014}.
This fact further complicates the bias correction methods for these experiments.

Here, we consider two different definitions of the anomaly, which differ by the choice of the reference period and data.
The first, and the most intuitive, approach is to consider the predictions without any bias corrections.
In order to easily visualize the data, a common factor (the same factor for all models and for all times) is subtracted from the predictions, that is:

\begin{equation}
 Y^c_{m,r}\left(t\right)= Y_{m,r}\left(t\right) - X\left(t\ mod \ 12\right),
\end{equation}
where $Y_{m,r}\left(t\right)$ is the prediction of the $r$th realization of model $m$ for time $t$ (in our analysis, $t$ measures a discrete number of months, and the variable $Y$ is either the surface temperature, $T$, or the surface zonal wind, $U$) and $X\left(i\right)$ is the climatological average of month $i$ ($i\in[0,11]$; $i=0$ corresponds to December and $i=1\ldots11$ correspond to January-November, respectively) of the climate variable $Y$. The climatology used in this work is based on the NCEP reanalysis data \cite{kalnay_ncep/ncar_1996} for the period 1976-2006. In what follows, we refer to this definition as the $clm$ anomaly and use the superscript $c$ to denote it.

\begin{figure}
\centerline{\includegraphics[width=0.5\linewidth]{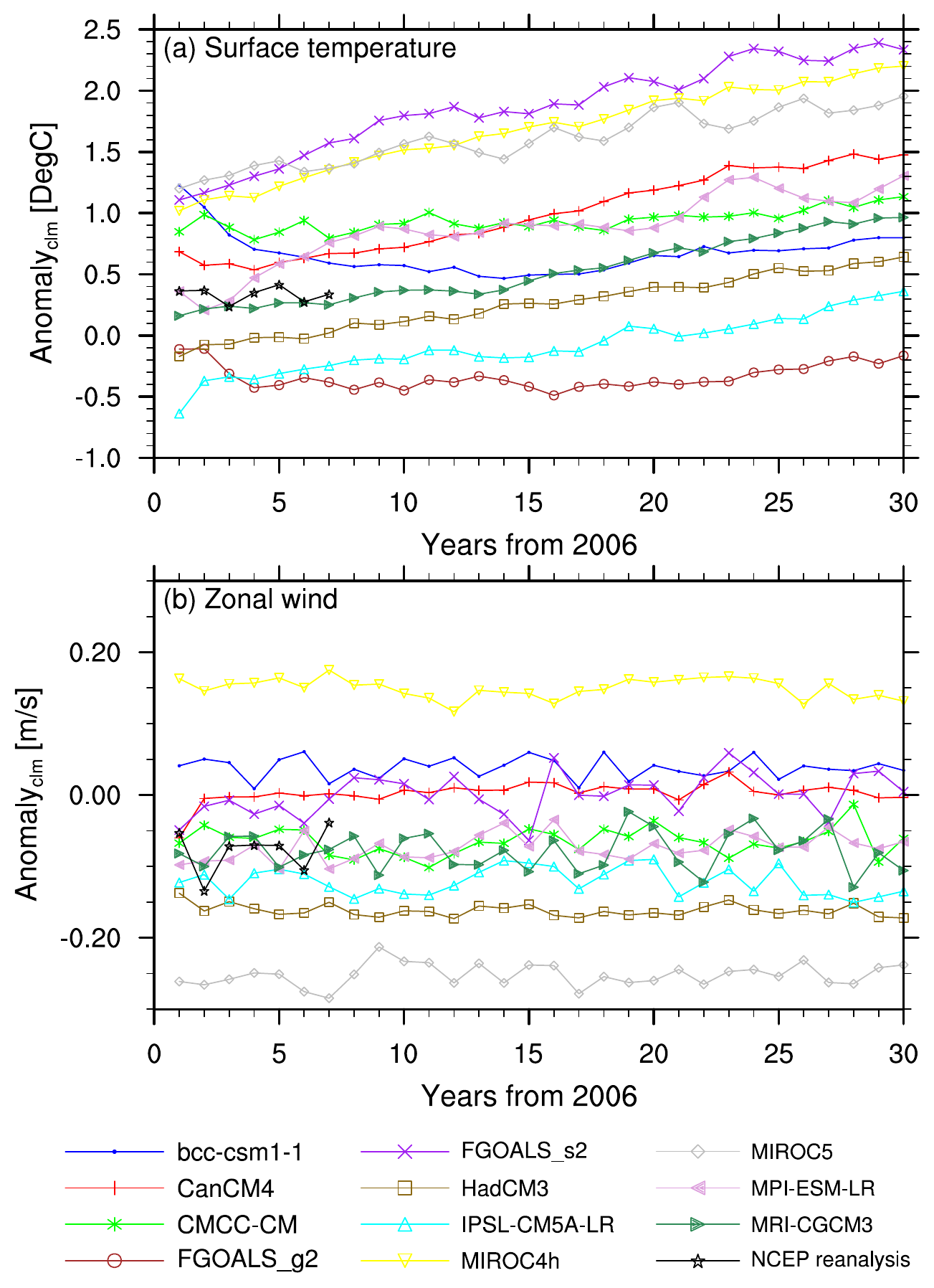}}
 \caption{Global averages of (a) $T^c_t$ and (b) $U^c_t$ as predicted by the models of the ensemble (averaged over the realizations). The values represent the predicted values minus a reference climatology derived from the NCEP reanalysis data for the period of 1976-2006. The factor subtracted is identical for all the models and, therefore, does not represent a bias correction.}\label{ncor_DATA_year}
\end{figure}

Figure \ref{ncor_DATA_year} shows the annual and global averages of $T^c_t$ and $U^c_t$ as predicted by the climate models in the ensemble (averaged over the realizations).
The NCEP reanalysis data for the first seven years is also presented for reference. Note that this is not a bias correction method since the same constant factor is removed from all the models.

As can be seen in the figure, most of the models predict an increase of $T^c$ with time.
Some of the models do show a decrease of $T^c$ during the first few years; this decrease can be associated with the drift from the initial condition to the model physics.
The spread of the model predictions also seems to increase with time but not substantially (see also the left column in Table \ref{trend and spread}).
The zonal surface wind, $U^c$, does not show any apparent trend in the data or in the spread of the model predictions.
In addition, none of the models show a drift toward different physics, except for the CAN-CM4 model, which predicts an increase of the zonal wind during the second year of the simulation and then smaller fluctuations around that value.

\begin{table}
\caption{The trends of the ensemble average and the spread for surface temperature ($T$) and surface zonal wind ($U$).
The trends are provided for the two definitions of the anomaly considered--no corrections anomaly (clm) and the bias correction anomaly (bias).
Note that by definition, the trend of the ensemble average is the same for the two definitions of the anomaly.}
\centering
\begin{tabular}{cccc}
\hline\noalign{\smallskip}
\tiny
 & Variable & clm & bias \\
\noalign{\smallskip}\hline\noalign{\smallskip}
\multirow{2}{80pt}{Trend of ensemble mean} & $T$ $[{}^\circ C y^{-1}]$ & $0.023$ & $0.023$ \\
\\ \cline{2-4}\\
                                        & $U$ $[ms^{-1}y^{-1}]$ & $2.4\times 10^{-3}$ & $2.4\times 10^{-3}$ \\
\noalign{\smallskip}\hline\noalign{\smallskip}
\multirow{2}{80pt}{Trend of model spread}  & $T$ $[{}^\circ C y^{-1}]$ & $0.0069$ & $0.0031$ \\
\\ \cline{2-4}\\
                                        & $U$ $[ms^{-1}y^{-1}]$ & $ 8.3\times 10^{-6}$ & $-1.8\times 10^{-5}$ \\
\noalign{\smallskip}\hline
\end{tabular}
\label{trend and spread}
\end{table}
\begin{figure}[!ht]
 \centerline{\includegraphics[width=0.5\linewidth]{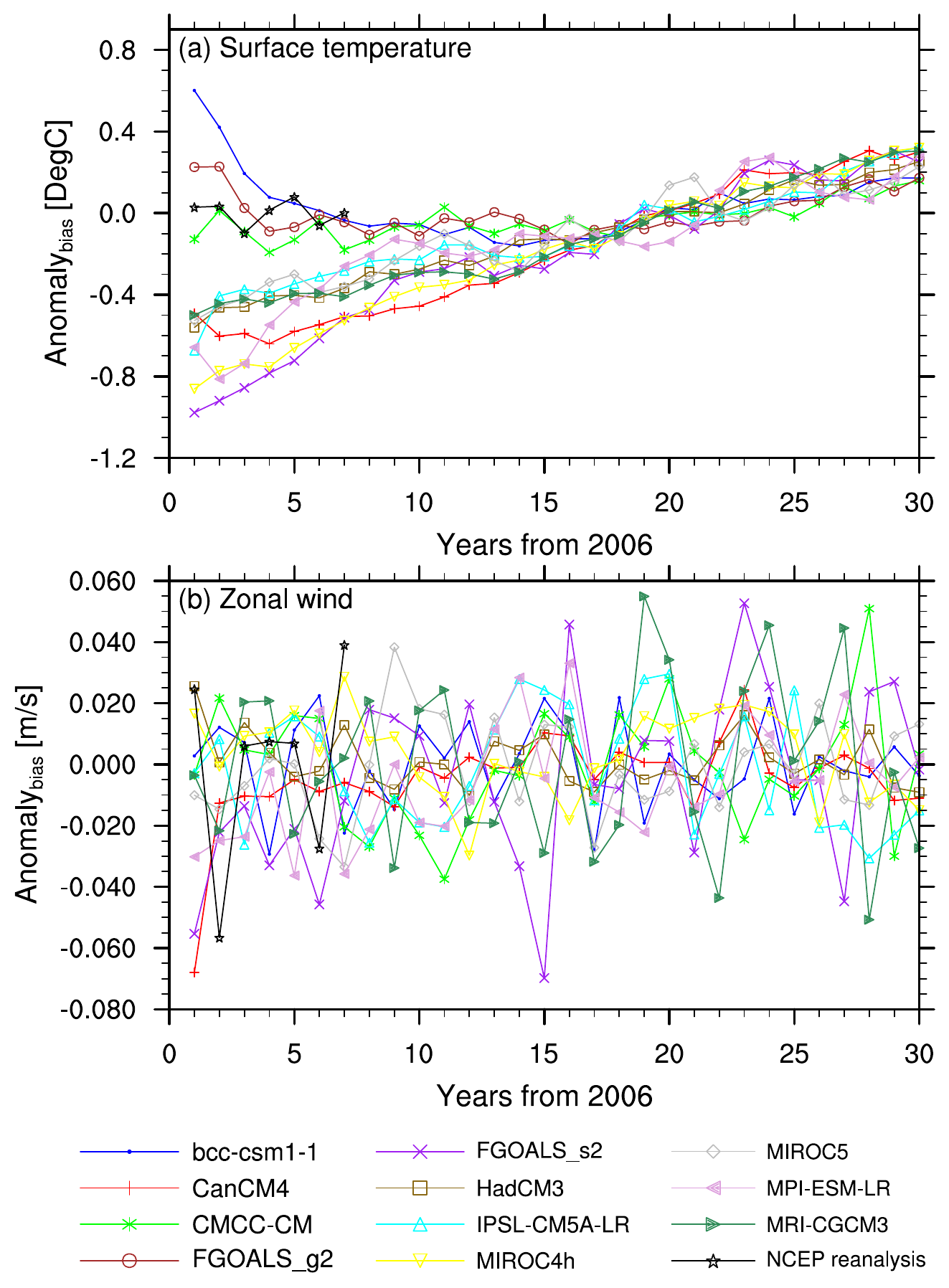}}
 \caption{Global averages of (a) $T^b_t$ and (b) $U^b_t$ as predicted by the models of the ensemble (averaged over the realizations).
The curves present the predicted values minus the 2016-2036 climatology of each model.
The factor subtracted is different for each model and represents a bias correction.
}\label{bias_DATA_year}
\end{figure}
The second definition of the anomaly refers to the deviation of each model from its own climatology during some reference period.
This definition removes the problems due to known biases of the models.
However, the predictions are only meaningful if one assumes that other characteristics of the climate variables (e.g., higher moments and extreme events) are not altered by the bias.
Here, the reference period for calculating the climatology of each model was the last twenty years of the simulations, i.e., 2016-2036.
This choice of the reference period is expected to remove the effects of the drift from the initial condition toward the model physics \cite{meehl_decadal_2014}.
We refer to this anomaly definition as the $bias$ anomaly and denote it with the $b$ superscript; the definition reads:
\begin{equation}
 Y^b_{m,r}\left(t\right)= Y_{m,r}\left(t\right) - \overline{ Y_{m,r}\left( t\ mod \ 12\right) }.
\end{equation}
Here, $\overline{Y_{m,r}\left(i\right)}$ is the 2016-2036 average during month $i$ ($i\in[0,11]$; $i=0$ corresponds to December and $i=1\ldots11$ correspond to January-November, respectively) of the climate variable $Y$ as predicted by the realization $r$ of model $m$.
This definition does not rely on the assumption that the quality of the prediction depends mainly on the lead time \cite{GRL:GRL29181} and is similar to the bias correction suggested in \cite{Goddard_2013} for anomaly-initialized models (see also \cite{meehl_decadal_2014}).

The $bias$ anomaly of the models is plotted in Figure \ref{bias_DATA_year} for $T^b$ and $U^b$. The $T^b$ panel shows that this definition of the anomaly highlights the effects of the model drift. During the first few years, some of the models drift to their preferred physical state, and then, they all predict an increase in the global temperature.
$U^b$, just like $U^c$, does not show a considerable trend (see the right column in Table \ref{trend and spread}).
Following the drift period, the spread of the model anomalies is much smaller than the spread of the $clm$ anomalies. This is not surprising because the $bias$ anomaly removes from each model its average, thereby bringing the model anomalies closer to each other.

We would like to mention two points in favor of the first definition of the anomaly ($clm$) that does not involve a bias correction.
Uncertainties and predictions derived from the $clm$ anomaly do not rely on the assumption that the characteristics of the climate variables are not altered by the biased climate simulated by the models. Clearly, the bias correction methods only correct the first moment (the mean) of the variable. Higher moments are likely to be altered in the simulated biased climate.
In addition, the $clm$ anomaly provides an equally good estimate of the uncertainty at all times and does not favor the period long after the initialization of the models.
A point in favor of the the second method ($bias$) is the forecast reliability. A quick look at Figure \ref{ncor_DATA_year} reveals that the variability of the models is much larger than the mean square error of the ensemble mean. For the $bias$ anomaly, we can see in Figure \ref{bias_DATA_year} that after the drift of the models during the first years of the simulations, the mean square error is close to the variability of the models, thereby allowing us to consider the ensemble mean of this anomaly as a reliable forecast \cite{palmer_2006}.
Since there is no straightforward solution to the question regarding which of the definitions of the anomaly is better, we will present the uncertainties for both anomalies.

\section{Global Properties of the Variability}

\subsection{Dependence of the variability on the averaging period}

\begin{figure}
 \centerline{\includegraphics[width=0.5\linewidth]{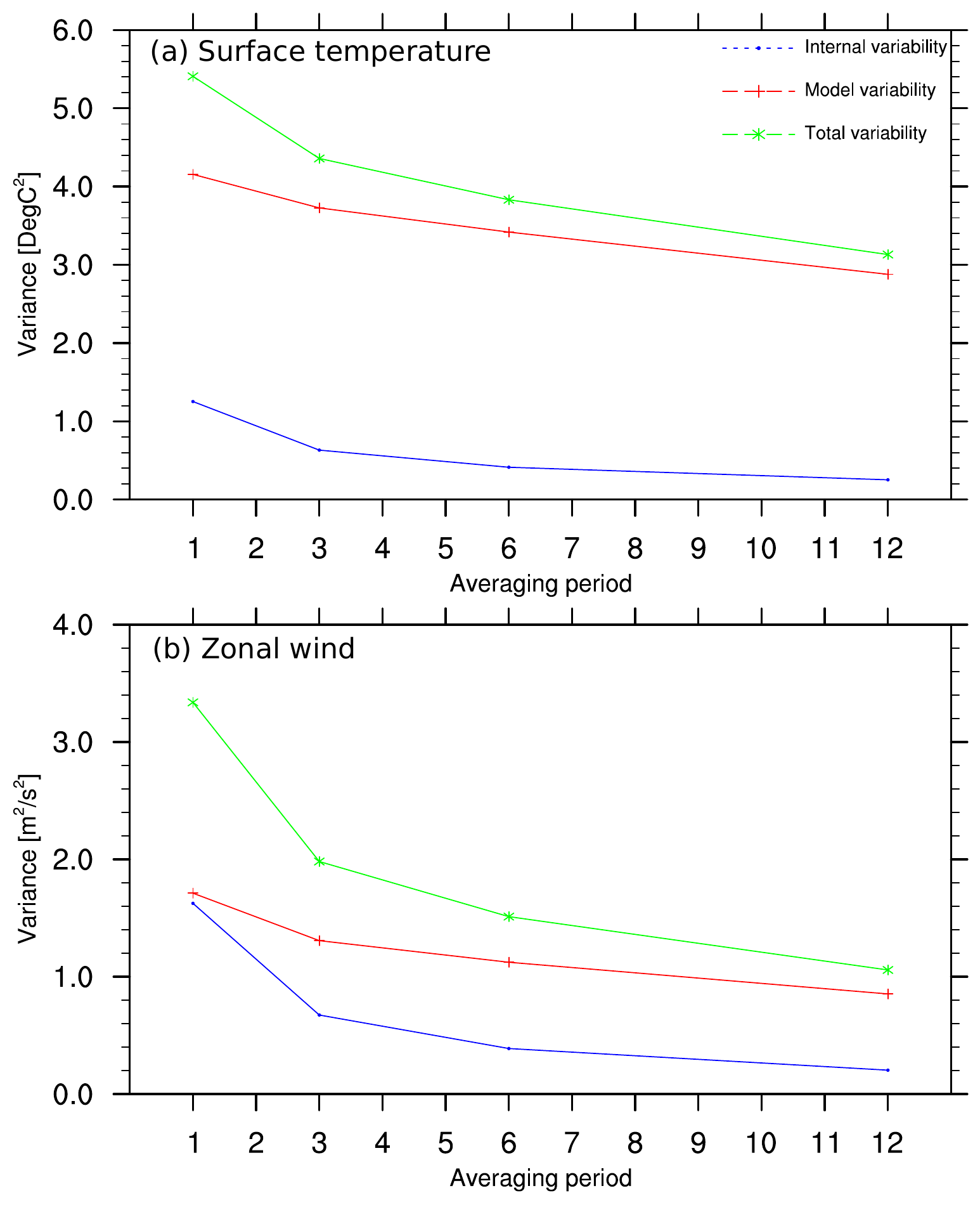}}
 \caption{Temporal and global averages of the total, model and internal variabilities of $T^c$ (panel a) and $U^c$ (panel b) (the $clm$ anomalies of the surface temperature and zonal wind speed, respectively), for four different averaging periods (1, 3, 6 and 12 months). The temporal averages were calculated from the 30-year prediction period.}\label{averaging_period}
\end{figure}

Climate simulations are only expected to provide meaningful predictions for the temporal averages of the climate variables. In decadal climate predictions, the period for temporal averaging varies from one month to several years. The uncertainty in the prediction of the climate variables strongly depends on the averaging period. This dependence is illustrated in Figure \ref{averaging_period}. In this figure, the 30-year averages of the total, model and internal variabilities are presented for four different averaging periods: 1, 3, 6, and 12 months. The results in Figure \ref{averaging_period} are for the $clm$ anomalies, $T^c$ and $U^c$.

As expected, the variability decreases for a longer averaging period. The sharpest decline occurs in the transition from the monthly to the seasonal averages.
The decrease in the variability of the surface zonal wind (panel b of Figure \ref{averaging_period}) is more significant than the decrease in the variability of the surface temperature.
The figure also shows that the sharp decrease in the variability is mainly due to the decrease of the internal variability.
For the $bias$ anomalies, the same trends are observed (not presented here), but the decrease in uncertainty is more significant because the internal variability constitutes a larger portion of the total uncertainty.

\subsection{Yearly averages of the variabilities}

\begin{figure}
 \centerline{\includegraphics[width=\linewidth]{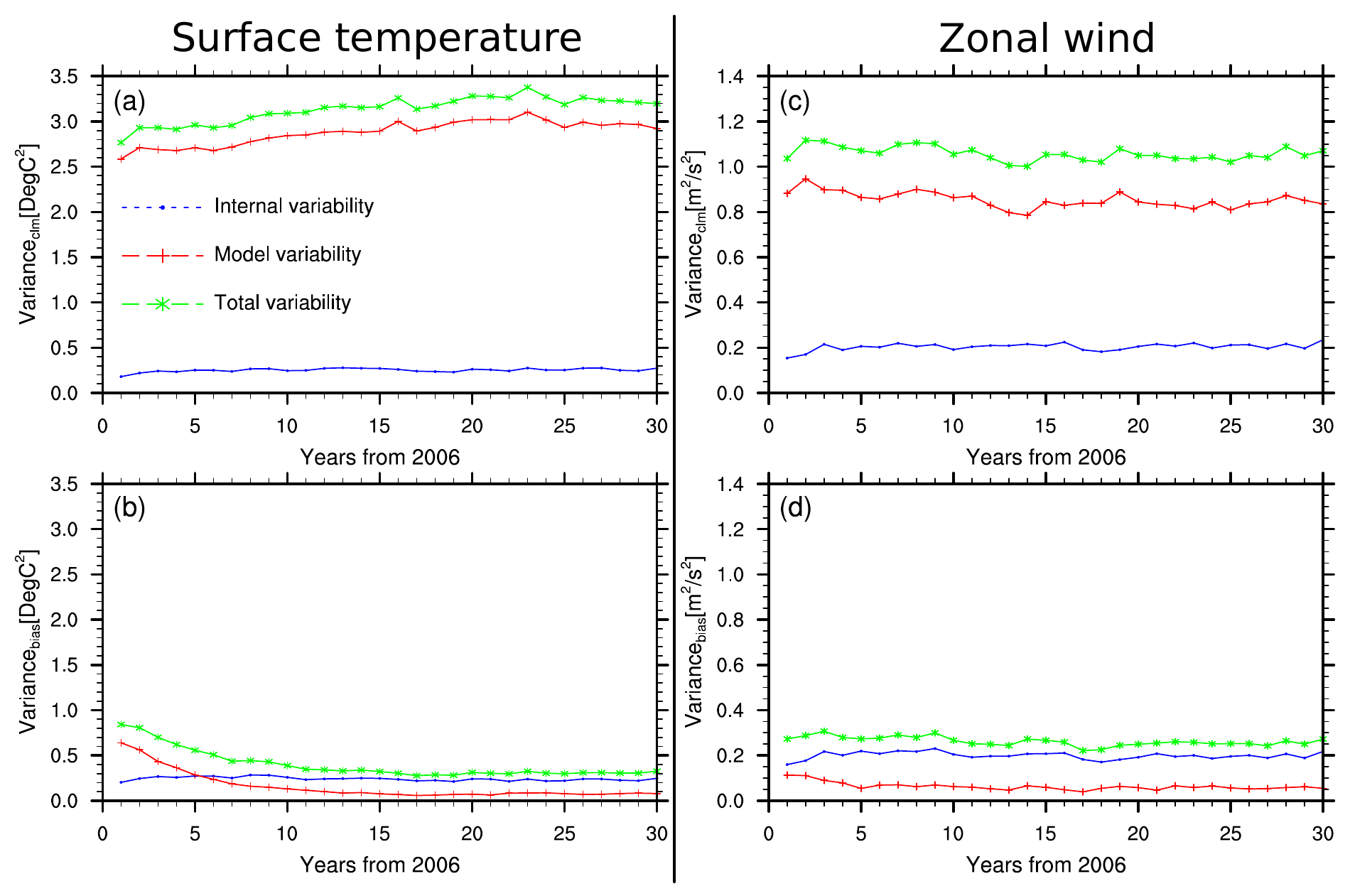}}
  \caption{Global averages of the internal, model and total variabilities of the annual means of surface temperature, $T$ (left) and zonal wind $U$ (right) anomalies. Panels (a,c) present the variabilities of the $clm$ anomaly and panels (b,d) present the variabilities of the $bias$ anomaly.}\label{VAR_year}
\end{figure}

\begin{figure}
 \centerline{\includegraphics[width=\linewidth]{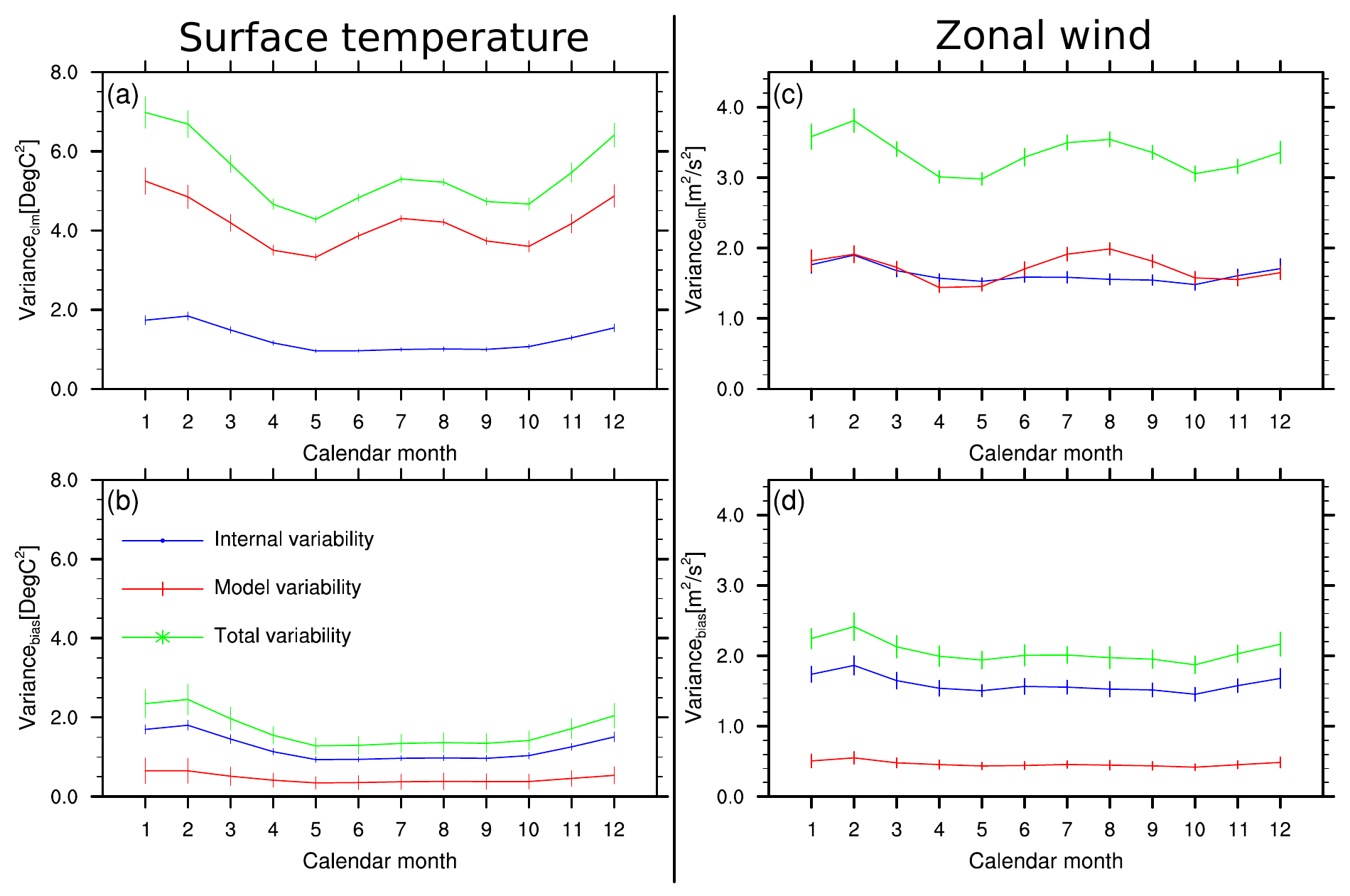}}
  \caption{Global and temporal averages of the internal, model and total variabilities of the monthly means of surface temperature, $T$ (left) and zonal wind $U$ (right) anomalies. The temporal average of each calendar month was calculated from the 30-year prediction period. Panels (a,c) present the variabilities of the $clm$ anomaly and panels (b,d) present the variabilities of the $bias$ anomaly.
The error bars represent two standard deviations calculated from the 30-year time series of the global average variance for each month.
}\label{VAR_month}
\end{figure}

The evolution of the uncertainties is of great importance. Naively, one would expect that the variability will grow with the lead time (the time since the initialization of the models).
In order to study this evolution, we present, in Figure \ref{VAR_year}, the global average of the variabilities of the annual mean anomalies of the surface temperature (left) and zonal wind (right) for the two anomaly definitions.
The upper panels present the variabilities of the $clm$ anomalies, and the lower panels show the variabilities of the $bias$ anomalies.
It is shown that the model variability is the main source of uncertainty in the $clm$ anomalies, while the internal variability is responsible for most of the uncertainty in the $bias$ anomalies.
In addition, we found that the model variability of $T^c$ shows a moderate increase with time, unlike its internal variability, which is almost constant during the prediction period.
The variabilities of $T^b$ show no clear trend during the prediction period except for the decrease of the model variability during the drift of the models from the initial condition to their preferred climate.
For the surface zonal wind, we found that both the internal and the model variabilities show no significant trend for either anomaly ($U^c$ or $U^b$).
Comparing the variabilities of the two anomalies (upper and lower panels), we found that the internal variability of the $bias$ anomalies is about the same as that of the $clm$ anomalies.
However, the model variability of the $bias$ anomaly is much smaller than that of the $clm$ anomaly. The smaller model variability in the $bias$ anomaly is expected because of the bias correction process.

\subsection{Intra-annual dynamics of the variability \label{sec_intraannual}}

The previous subsections focused on the variabilities of the annual means and their evolution.
However, it is natural to expect that different months will have different contributions to the variabilities.
We considered the monthly means of the surface temperature and zonal wind anomalies in order to investigate the intra-annual dynamics of the variabilities.
The variabilities were averaged over the whole globe (the weight of each grid cell is proportional to the area it spans) and over the 30 years of the prediction period.
Figure \ref{VAR_month} shows the global and 30-year average variabilities of the $T$ and $U$ anomalies for each of the calendar months.
The error bars represent two standard deviations calculated from the 30-year time series of the global average variance for each month.
The spatial variability of the variance for each month is presented in the Supplementary Materials.
As expected, the variability of the monthly means is higher than the variability of the annual means.
For example, the internal variability of the monthly means of $T^c$ is in the range of $1-2 \ \ ^{\circ} C^2$, whereas the internal variability of the annual means of $T^c$ is in the range of $0.1-0.3 \ \ ^{\circ} C^2$--a reduction of an order of magnitude. A smaller reduction is found in the model variabilities.
These findings indicate that we might predict the annual averages of a climate variable with a relatively small uncertainty, but this is only because the shorter time scale fluctuations are averaged out.
The internal variability shows one significant peak during the northern hemisphere winter, whereas the model variability of the $clm$ anomalies has two peaks: one (larger) in the
northern hemisphere winter and one (smaller) in the northern hemisphere summer.
One can also see that the $bias$ anomalies have much smaller variabilities compared with the $clm$ anomalies due to the bias correction.
The error bars suggest that the inter-annual fluctuations in the uncertainty are larger for the zonal wind than for the surface temperature anomalies.

\section{Regional Properties of the Variability}

\begin{figure}[!ht]
  \centerline{\includegraphics[width=0.5\linewidth]{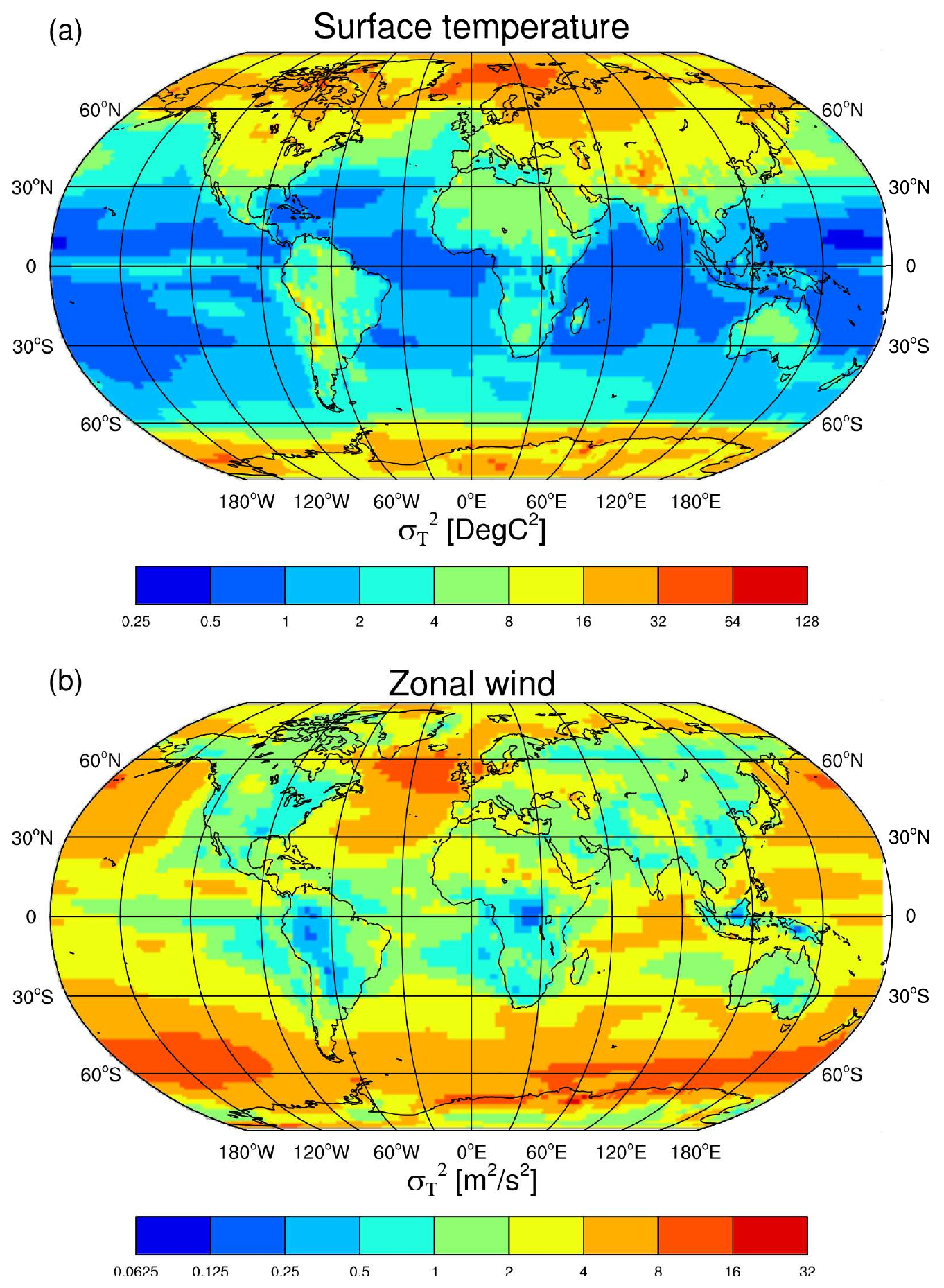}}
   \caption{Temporal average of the total variability, $\sigma_T^2$, (in log scale) of the annual mean $clm$ anomalies of (a) surface temperature, $T^c$, and (b) surface zonal wind, $U^c$. The temporal average was calculated from the 30-year prediction period.}\label{VAR_global_total}
 \end{figure}

The variability is not only affected by the averaging period and the season but it also strongly depends on the location and shows a large spatial variability.
Figure \ref{VAR_global_total} shows the spatial distribution of the total variability, $\sigma_T^2$, of the annual means of the $clm$ anomalies.
The variabilities presented correspond to the average variability during the 30-year prediction period.
The upper panel shows the total variability of the surface temperature anomaly, $T^c$, and the lower panel shows the total variability of the surface zonal wind anomaly, $U^c$.
The spatial variability is apparent for both climate variables.
The variability of $T^c$ is larger in higher latitudes and in coastal regions, whereas the $U^c$ shows high variability mainly over the oceans.
The topography also affects the variability, such as in the Andes and the Himalayas.

\begin{figure}[!ht]
 \centerline{\includegraphics[width=0.5\linewidth]{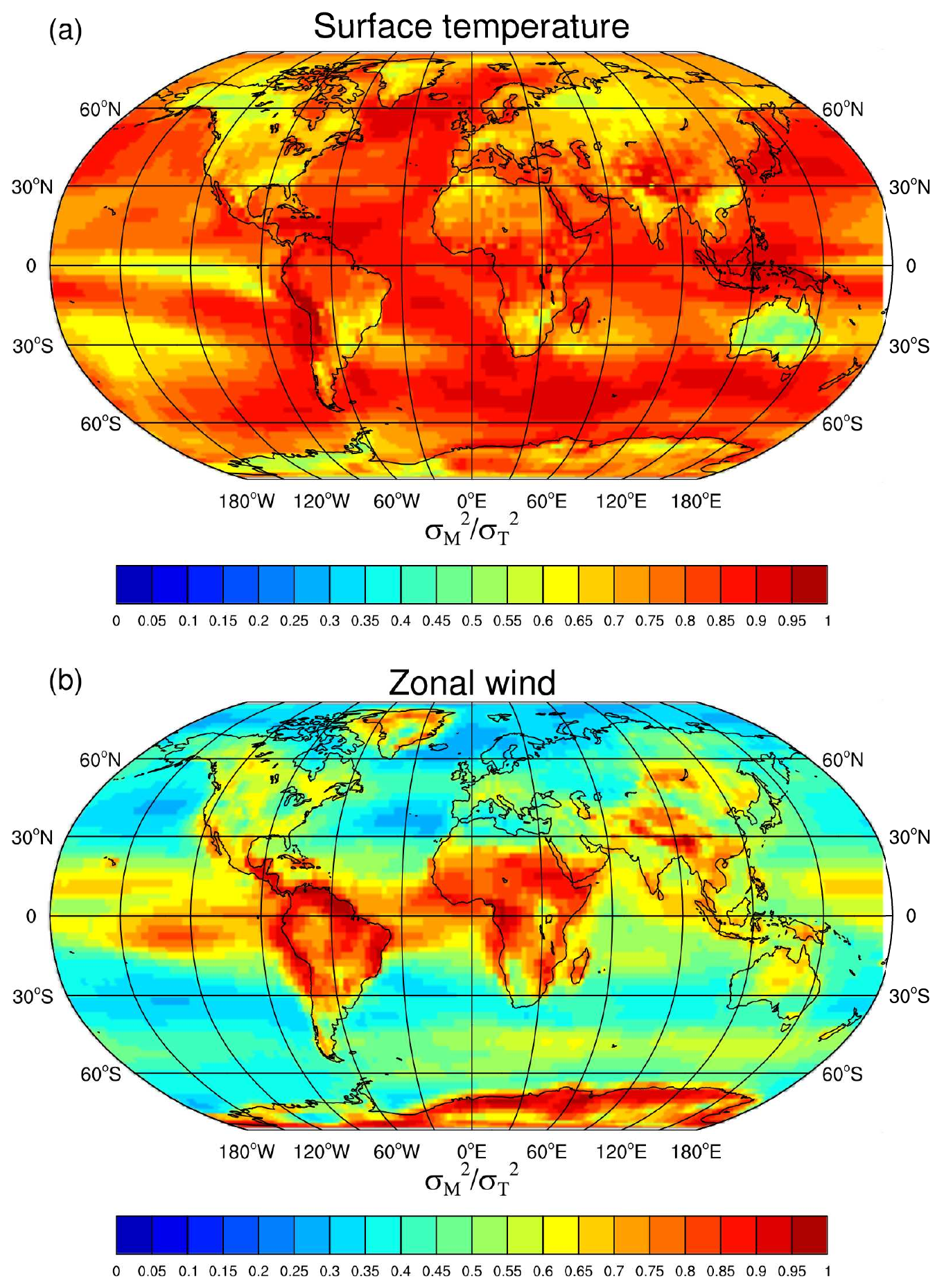}}
  \caption{The fraction of model variability from the total variability, $\sigma_M^2/\sigma_T^2$ , for (a) $T^c$, and (b) $U^c$.}\label{VAR_global_model_frac}
\end{figure}

To further analyze the role of the two main sources of variability (internal and model), we present, in Figure \ref{VAR_global_model_frac}, the spatial distribution of the fraction of the model variability, $\sigma_M^2/\sigma_T^2$.
The upper panel shows that the model variability constitutes the main part of the uncertainty associated with $T^c$ almost everywhere.
The lower panel shows that over land, the model variability is also the main source of $U^c$ uncertainty, while over the oceans, it is smaller than the internal variability.

\begin{figure}[!ht]
 \centerline{\includegraphics[width=\linewidth]{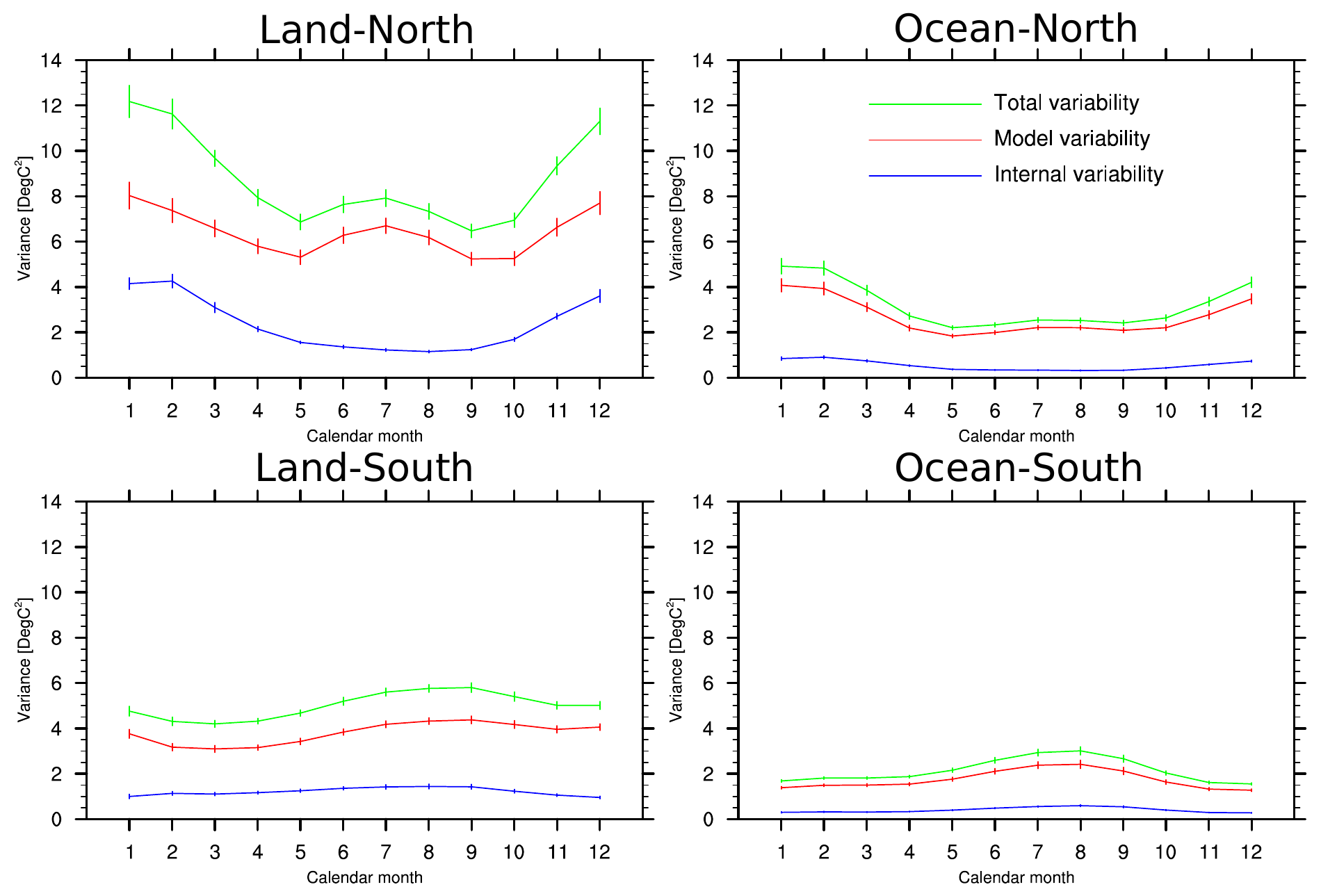}}
  \caption{Temporal average of the $T^c$ internal, model and total variabilities for four domains--over land in the northern hemisphere, over land in the southern hemisphere, over ocean in the northern hemisphere, over ocean in the southern hemisphere. The temporal averages were calculated from the 30-year prediction period for each calendar month. The error bars represent two standard deviations calculated from the 30-year time series of the global average variance for each month.}\label{VAR_tas_month}
\end{figure}

Motivated by the results presented in Figure \ref{VAR_global_total}, we plotted the monthly average internal, model and total variabilities for four regions over the globe: (a) Land-North, (b) Land-South, (c) Ocean-North and (d) Ocean-South.
The contributions of these regions to the global average variability (i.e., their relative surface area compared with the global surface area) are $19.5\%$, $9\%$, $30.5\%$ and $41\%$, respectively.
Figure \ref{VAR_tas_month} shows that the main contribution to the global variability of $T^c$ comes from the land areas in the northern hemisphere.
The contributions to the global averaged variability of $T^c$ of the northern hemisphere's oceans and the southern hemisphere's lands are both comparable (and smaller than the contribution of the Land- North region), and the contribution of the southern hemisphere's oceans, which constitute the largest region, is the smallest.

\begin{figure}[ht]
 \centerline{\includegraphics[width=\linewidth]{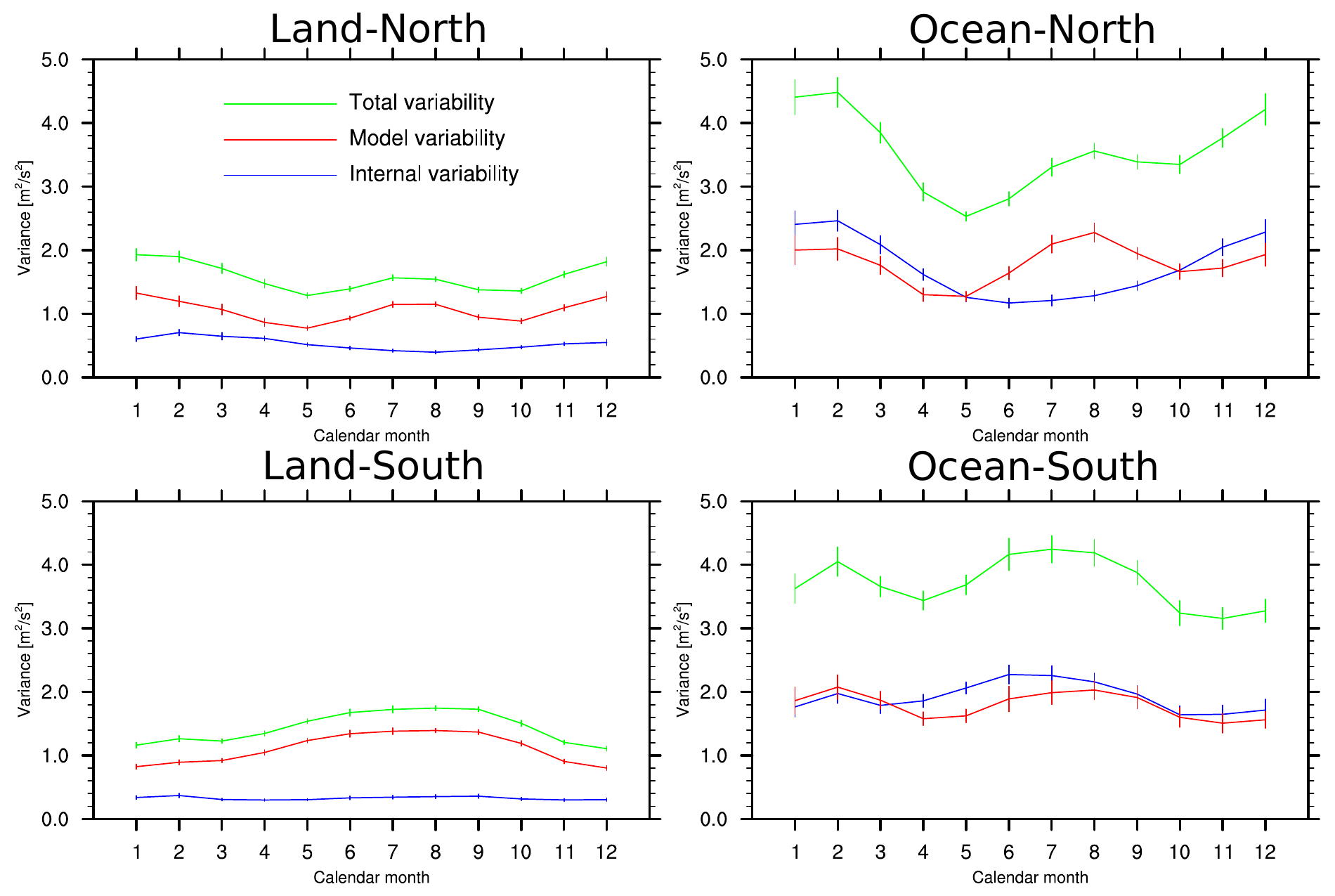}}
  \caption{Temporal average of the $U^c$ internal, model and total variabilities for four domains--over land in the northern hemisphere, over land in the southern hemisphere, over ocean in the northern hemisphere, over ocean in the southern hemisphere. The temporal averages were calculated from the 30-year prediction period for each calendar month. The error bars represent two standard deviations calculated from the 30-year time series of the global average variance for each month.}\label{VAR_uas_month}
\end{figure}

The $U^c$ variabilities are shown in Figure \ref{VAR_uas_month}.
For the surface zonal wind, the two hemispheres' oceans are the main sources of variability.
The variability over land has a smaller contribution to the globally averaged variability with the smallest contribution coming from the southern hemisphere's land areas.
The model variability is higher over land compared with the internal variability as can be seen also in Figure \ref{VAR_global_model_frac}(b).

Figures \ref{VAR_tas_month} and \ref{VAR_uas_month} indicate that the internal variability is highest in winter in both hemispheres.
However, the reason that only the peak during the winter in the northern hemisphere is visible in Figure \ref{VAR_month} is that the variability in the northern hemisphere winter is much larger than that during the southern hemisphere winter.
For the surface temperature, it is the northern hemisphere's land areas that control the profile of the annual cycle, while for the surface zonal wind, it is the balance between the northern and the southern hemispheres' oceans that dictates the profile of the annual cycle. The two peaks observed in the globally averaged model variability of the surface temperature (see Figure \ref{VAR_month}(a)) are due to land areas in the northern hemisphere, while the two peaks in the globally averaged model variability of the surface zonal wind are mainly due to the oceans in both hemispheres. These peaks during the winter and summer indicate that the models have a larger disagreement in predicting the extreme climate conditions in these seasons.
In the Supplementary Materials, we include full global variability maps for each of the months for in-depth analysis of the spatial distribution of the variability components.

\section{Summary and Discussion}

Meaningful climate predictions must be accompanied by uncertainties.
While long-term climate predictions focus on the response of the climate system to changes in the atmospheric composition, decadal climate predictions attempt to provide synchronized climate predictions for shorter periods. Therefore, the meaningful climate variables differ by the averaging period. For example, in decadal climate predictions, seasonal and annual means are expected to be synchronized with the actual dynamics, while for long-term climate predictions, only the decadal (or longer averaging periods) means are expected to be relevant. The main sources of uncertainty in decadal climate predictions are the model variability and the internal variability, reflecting the sensitivity of the models to the initialization method and the differences between parameterization schemes for processes that are not explicitly resolved by the models. A common practice in analyzing climate prediction is to apply ``bias correction'' procedures in order to overcome known tendencies of the models to simulate an average climate state, which is shifted from the observed one by a constant amount (often the global temperature is higher or lower than the measured one). However, the bias correction assumes that the climate dynamics predicted by the models is not affected by the bias. Moreover, this correction is applied as an empirical correction rather than relying on a physical basis. The variety of bias correction methods is the reason that a well-defined anomaly that is considered as the meaningful prediction of the models does not exist.

Here, we used an ensemble of the CMIP5 decadal predictions, focusing on two definitions of the anomaly. The $clm$ anomaly does not involve any bias correction, simply shifting the predictions to reflect the deviations from the NCEP reanalysis climatology of the 30 years prior to the simulated period. The $bias$ anomaly involves bias correction, reflecting the deviation of each model from its own climatology during the last 20 years of the prediction period. The choice of the model climatology only 10 years after the initialization was motivated by the observation that the first several years are strongly affected by the model's drift. In the Supplementary Materials, we also present the results for a third definition of the anomaly in which the deviations of each model from the climate it predicted for the year 2016 are considered.
The uncertainties were decomposed to those existing due to the models' sensitivity to initialization methods and initial conditions, the internal variability, and the variability between different models.
The definition of the variability ensured that the contributions of the internal and the model variabilities are independent and that the total ensemble variability is the sum of these two contributions.

We showed that the variability of the CMIP5 decadal predictions does not increase significantly with time.
For the $bias$ anomaly, most changes occur during the drift period, while for the $clm$ anomaly, there are no significant changes.
The results for the internal variability are in agreement with previous results for the long-term experiments of the CMIP3 \cite{hawkins_potential_2009,yip_simple_2011}.
However, in the decadal experiments of the CMIP5, we have not seen the increase in model variability that was found in the CMIP3 data \cite{hawkins_potential_2009,yip_simple_2011}.
The differences might be related to the different averaging periods (decadal in \cite{hawkins_potential_2009,yip_simple_2011} and, at most, annual in our work), different spatial resolutions (global and $15\degree \times 15\degree$ resolutions in \cite{yip_simple_2011,hawkins_potential_2009} and $2.5\degree \times 2.5\degree$ in our work) and, in particular, to the different nature of decadal climate predictions compared with long-term climate predictions.

The predictions and, in particular, the uncertainties strongly depend on the definition of the anomaly.
Bias correction reduces the model variability by subtracting a different climatology from each model.
In addition, the uncertainties also depend on the averaging period. The variability of the monthly means is much larger than that of the annual means.
The significant reduction in the variability is seen in the transition from monthly means to seasonal (3 months) means. The model variability is more sensitive to the annual cycle than the internal variability.
The relative importance of the model and internal variabilities depends on the definition of the anomaly; for the $clm$ anomalies, the model variability is larger than the internal variability and vice versa for the $bias$ anomaly. By analyzing the spatial distribution of the variabilities, we showed that the land areas in the northern hemisphere are the main source of uncertainty in surface temperature, while the oceans in both hemispheres are responsible for most of the variability of the surface zonal wind. Previous results \cite{hawkins_potential_2009} showed no clear distinction between the contributions of land and ocean areas but rather a latitude dependence of the variabilities. Some of the spatial patterns of the internal variability that we found for the surface temperature, such as the increased internal variability toward the poles, are in agreement with previous works \cite{hawkins_potential_2009}. However, the spatial patterns of the variabilities of the surface wind are different and were not presented in previous works.  

The results presented here suggest that the modeling of the climate dynamics in the mid-latitudes should be improved. Such an improvement is expected to reduce the uncertainty of surface temperature, which is mostly due to the mid-latitude land areas in the northern hemisphere. The predictions of the surface zonal wind over the oceans should be improved as well. Obviously, the uncertainty in winter is the largest due to storms and extreme climate conditions. A weighted ensemble of climate models may significantly reduce the uncertainties by weighting each of the models according to its past performance. The contribution of the model variability may be significantly reduced this way. However, a reduction of climate prediction uncertainties does not necessarily lead to improved forecast quality. A reduction of the uncertainties without any improvement of the predictions can lead to a situation in which all the models predict similar dynamics that do not span the measured state of the climate system, and thus, the ensemble is not reliable (or is overconfident). A reduction of climate variability must be associated with an improvement in climate predictions.

\begin{acknowledgments}
The research leading to these
results has received funding from the European Union Seventh
Framework Programme (FP7/2007-2013) under grant number
[293825].
\end{acknowledgments}

\bibliographystyle{unsrt}
\bibliography{references}

\appendix
\section{Alternative definition of the anomaly--deviations from the 2016 climatology}

In the main text, we showed that the variabilities of the decadal climate predictions strongly depend on the definition of the anomalies.
Here, we present the analysis of the variabilities of an additional definition. Here, the anomaly is defined to be the deviation from the climatology of 2016. The idea behind choosing 2016 is that 10 years after the initialization of the models, the effects of the drift are expected to be negligible.
This anomaly is defined as:
\begin{equation}
 Y^s_{m,r}\left(t\right)= Y_{m,r}\left(t\right) - Z_{m,r}\left(t\ mod \ 12\right),
\end{equation}
where $Y_{m,r}\left(t\right)$ is the prediction of the $r$th realization of model $m$ for time $t$ (in our analysis, $t$ measures a discrete number of months, and the variable $Y$ is either the surface temperature, $T$, or the surface zonal wind, $U$) and $Z_{m,r}\left(i\right)$ is the monthly mean of the variable $Y$ during month $i$ of 2016 ($i\in[0,11]$; $i=0$ corresponds to December and $i=1\ldots11$ correspond to January-November, respectively) as predicted by the realization $r$ of model $m$. In what follows, we refer to this definition as the $2016$ anomaly and use the superscript $s$ to denote it.

\begin{figure}
 \centerline{\includegraphics[width=19pc]{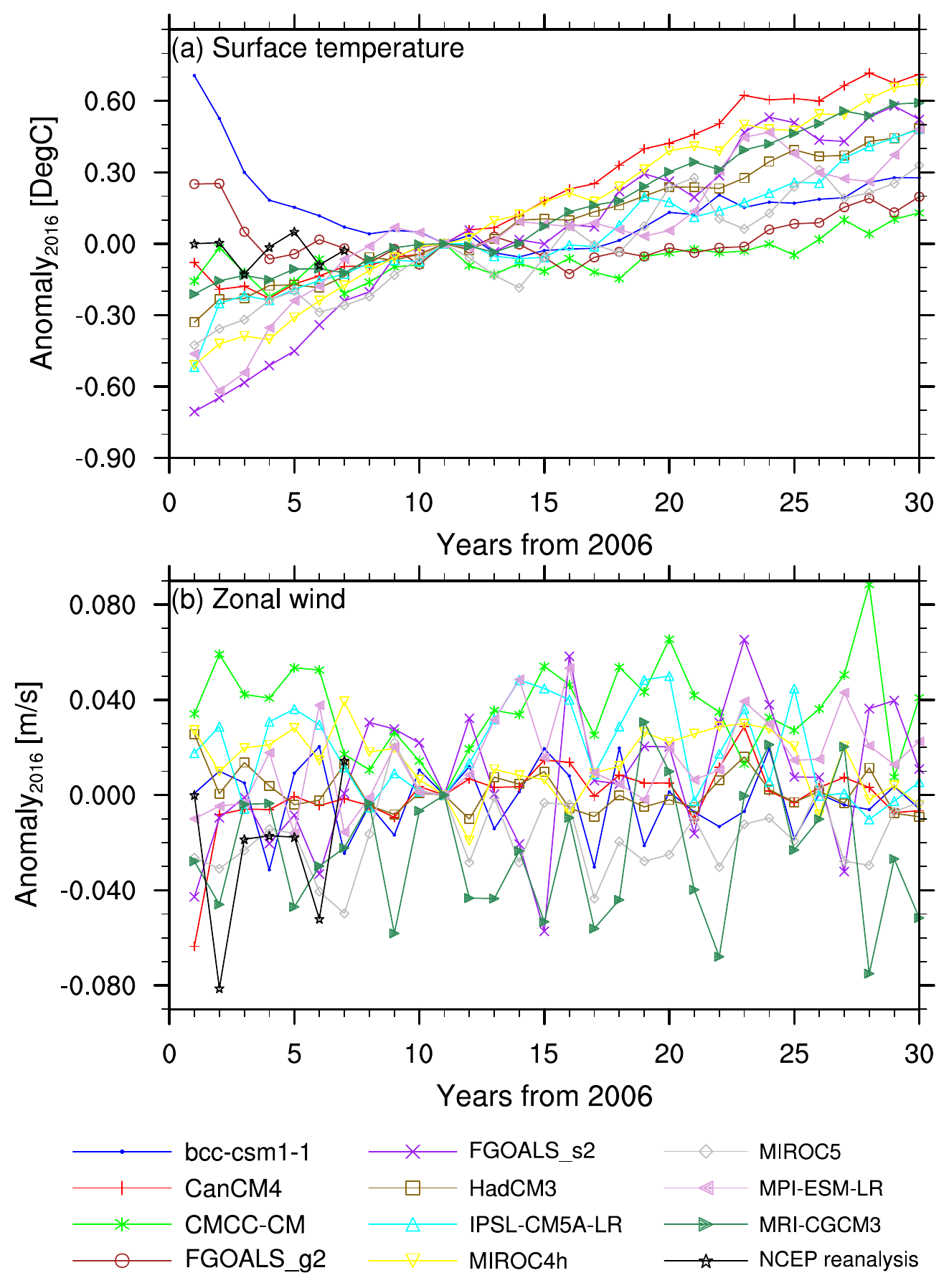}}
 \caption{Global and annual averages of (a) $T^s_t$ and (b) $U^s_t$ as predicted by the models of the ensemble (each model is averaged over its realizations).
 The curves present the predicted values minus the 2016 predictions of each model. The factor subtracted is different for each model and represents a bias correction. Surface temperature anomalies have a large spread between models during the first 10 years associated with the drift of the models to their climatology. In 2016, by definition, all the anomalies are zero. The spread between models after the first 10 years is smaller and increases with time. The surface zonal wind anomalies show no considerable change between the first 10 years of the predictions and the last 20 years.}\label{vas_DATA_year}
\end{figure}

\begin{figure}
 \centerline{\includegraphics[width=39pc]{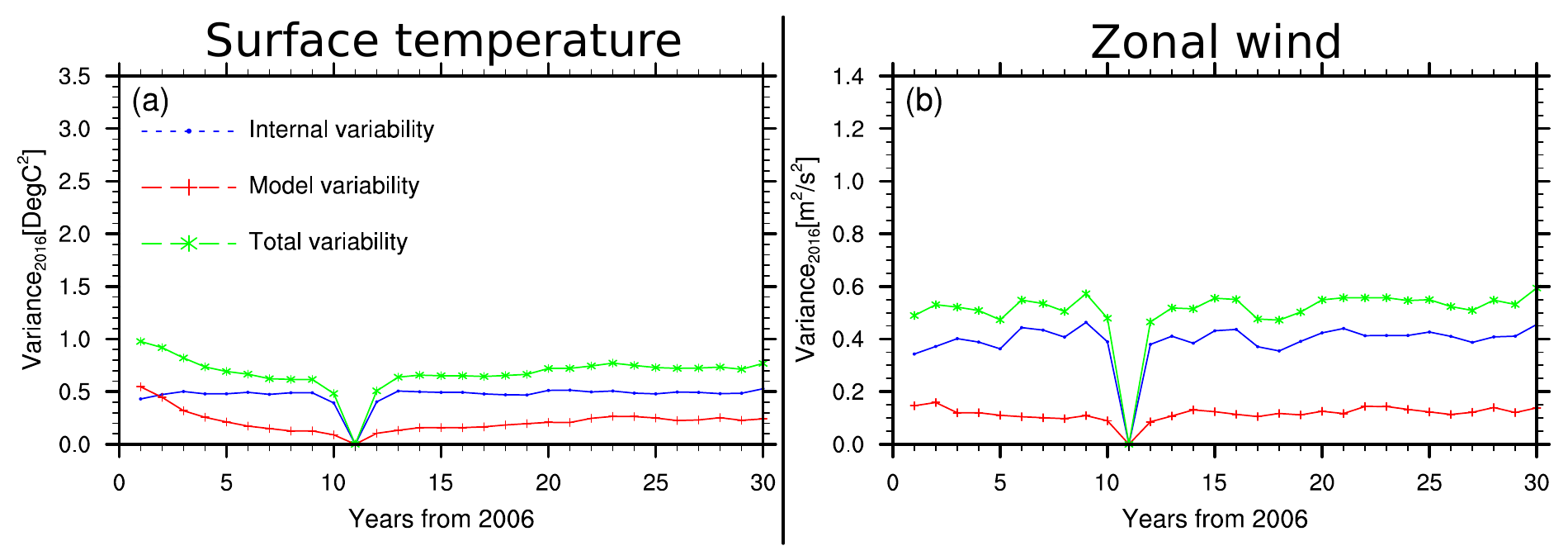}}
 \caption{Global averages of the internal, model and total variabilities of the annual means of the surface temperature, $T^s$ (left) and zonal wind $U^s$ (right) anomalies.
 Both panels present the variabilities of the $2016$ anomaly. There are no considerable trends in the variability that can be observed, except for a small trend during the first 10 years of the predictions, which could be associated with the drift of the models to their climatology. }\label{VAR_year_zero}
\end{figure}

\begin{figure}
 \centerline{\includegraphics[width=39pc]{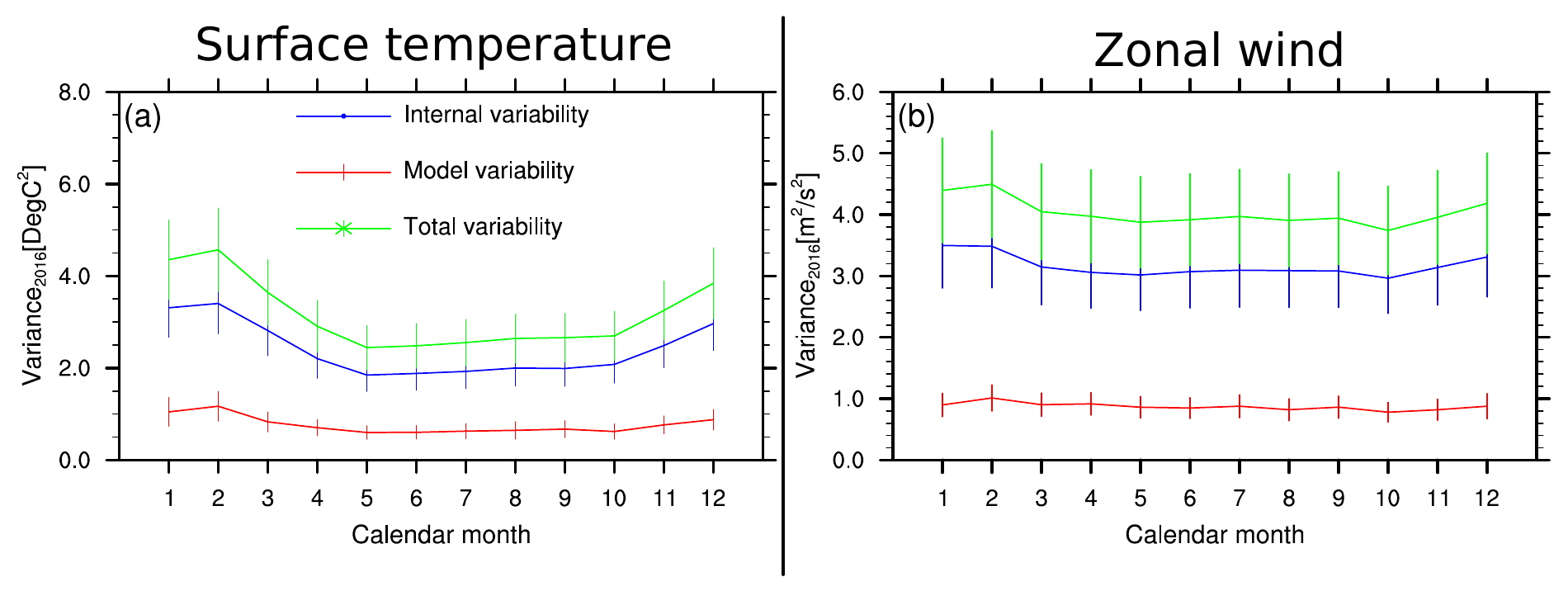}}
 \caption{Global and temporal averages of the internal, model and total variabilities of the monthly means of surface temperature, $T^s$ (left) and zonal wind $U^s$ (right) anomalies. The temporal average of each calendar month was calculated from the 30-year prediction period. Both panels present the variabilities of the $2016$ anomaly. The error bars represent two standard deviations calculated from the 30-year time series of the global average variance for each month. The variabilities have one annual peak in the northern hemisphere winter (for the zonal wind, the peak is fairly small). }\label{VAR_month_zero}
\end{figure}

The time series of the surface temperature and zonal wind anomalies are presented in Figure \ref{vas_DATA_year}. The variabilities of the annual means of the $2016$ anomalies are presented in Figure \ref{VAR_year_zero}. Figure \ref{VAR_month_zero} depicts the variabilities of the monthly means. The variability of each calendar month is averaged over the 30 years of the prediction period, and the error bars represent two standard deviations.

\section{Meridional wind anomalies and their variabilities}

\begin{figure}[ht]
 \centerline{\includegraphics[width=19pc]{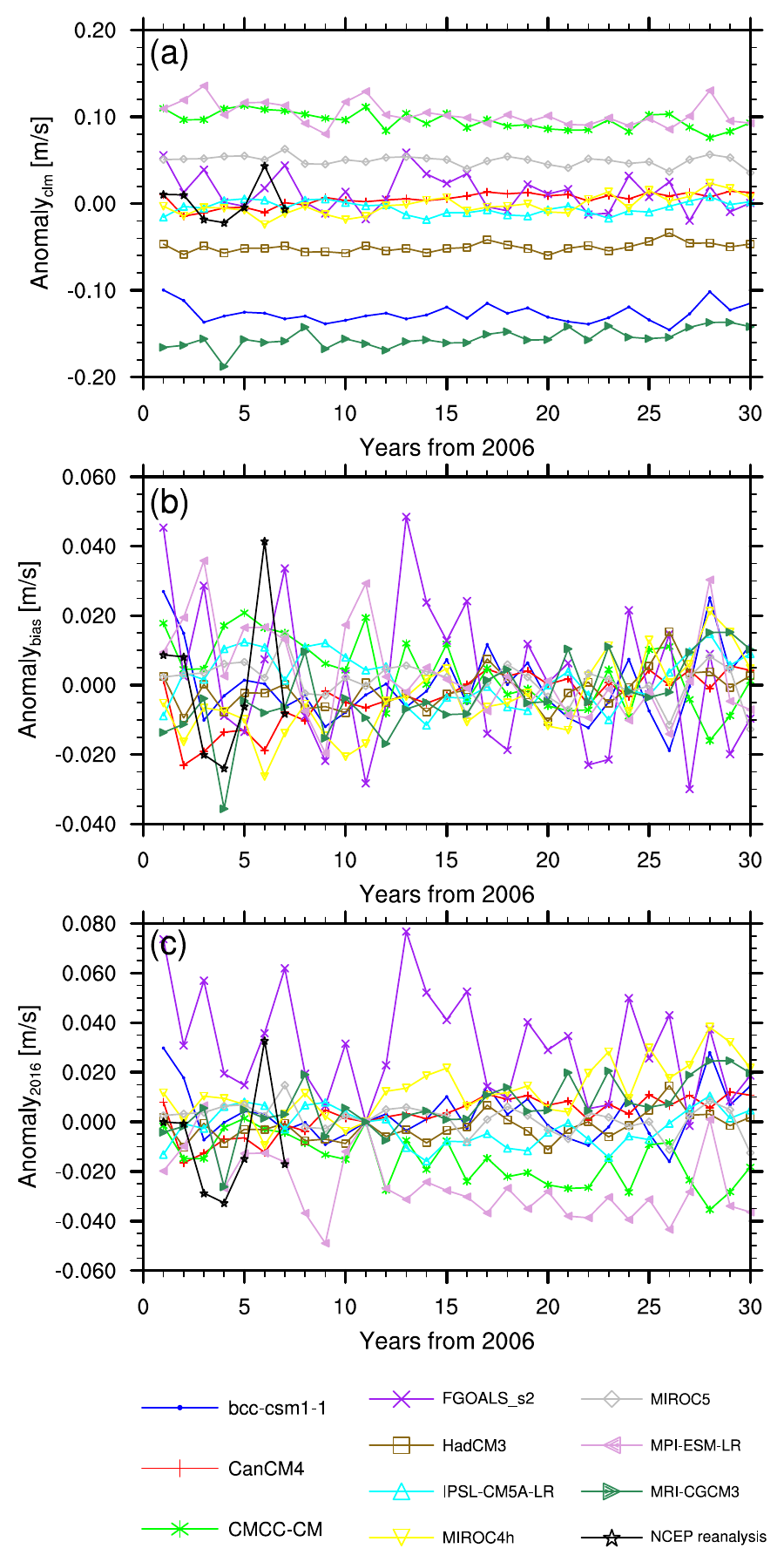}}
 \caption{Global and annual averages of $V_t$ as predicted by the models of the ensemble (each model is averaged over its realizations).
 The curves present the predicted values based on the (a) $clm$, (b) $bias$, and (c) $2016$ anomalies. No significant trend is observed.}\label{DATA_vas_year}
\end{figure}

\begin{figure}[ht]
 \centerline{\includegraphics[width=19pc]{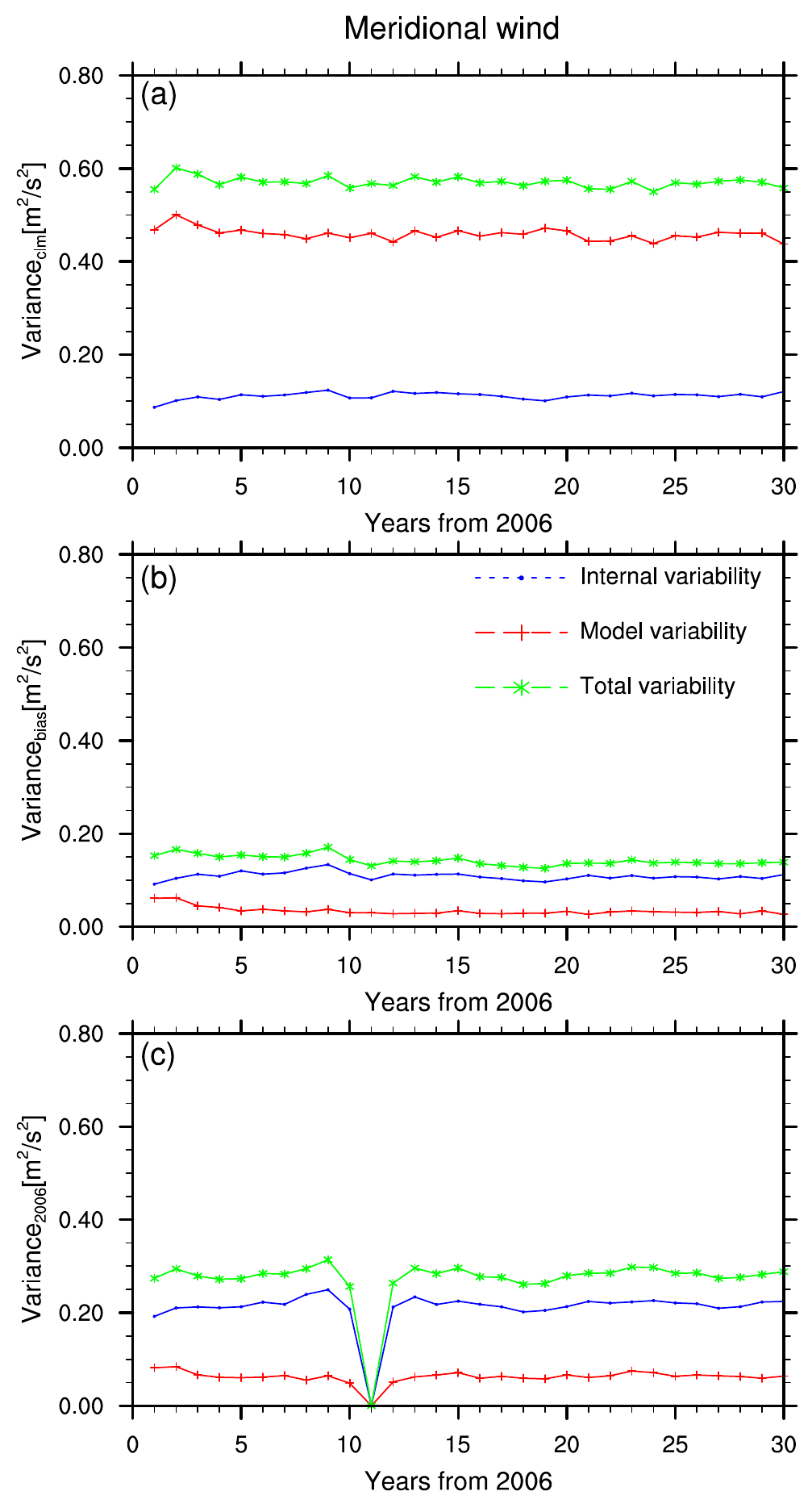}}
  \caption{Global averages of the internal, model and total variabilities of the annual means of (a) $clm$, (b) $bias$, and (c) $2016$ surface meridional wind anomalies.
  Both bias corrections (the $bias$ and $2016$ anomalies) significantly decrease the model variability.}\label{VAR_year_vas}
\end{figure}

\begin{figure}[ht]
 \centerline{\includegraphics[width=19pc]{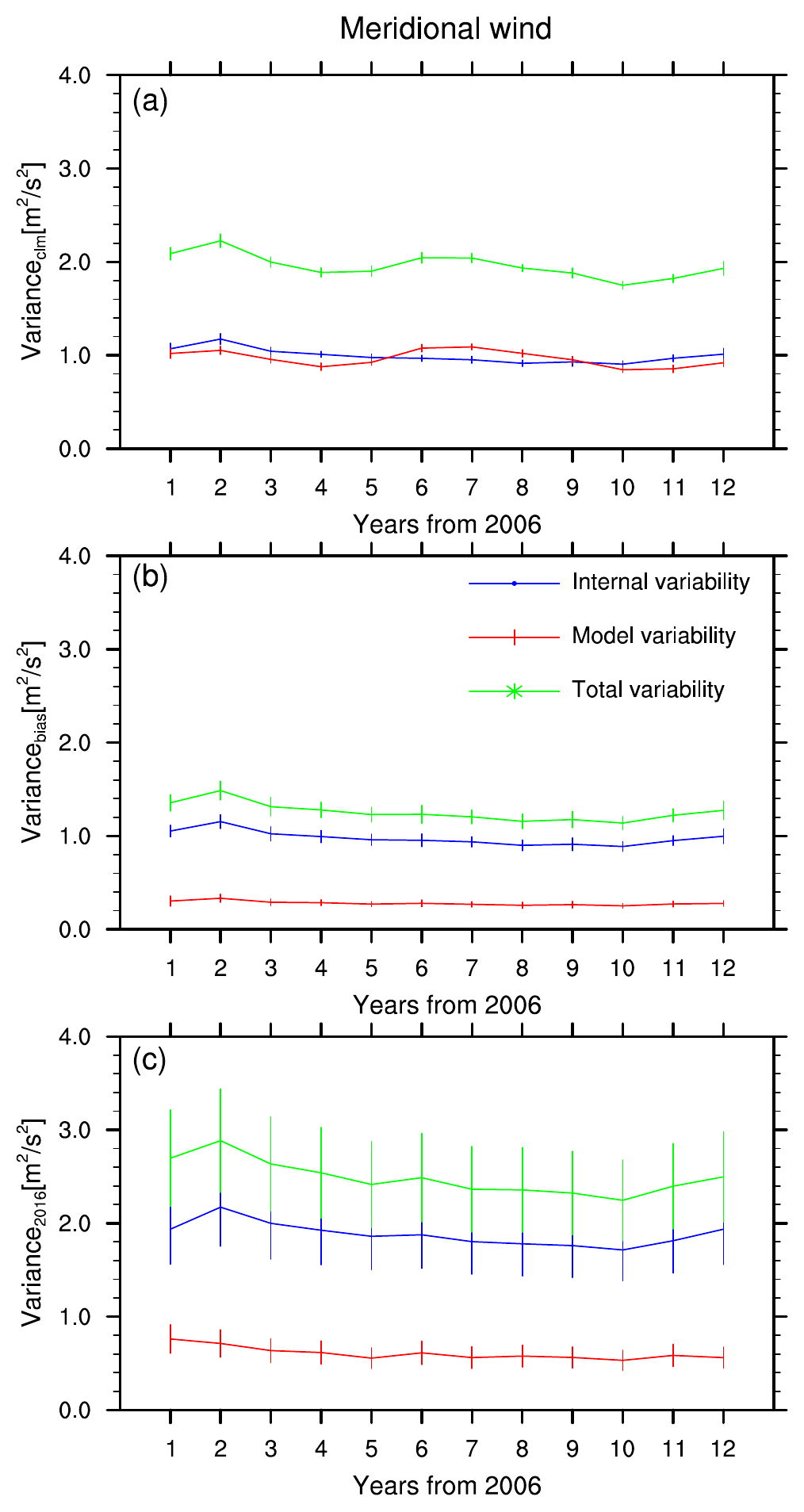}}
  \caption{Global and temporal averages of the internal, model and total variabilities of the monthly means of the surface meridional wind anomalies, (a) $clm$, (b) $bias$ and (c) $2016$.}\label{VAR_month_vas}
\end{figure}

\begin{figure}[ht]
 \centerline{\includegraphics[width=19pc]{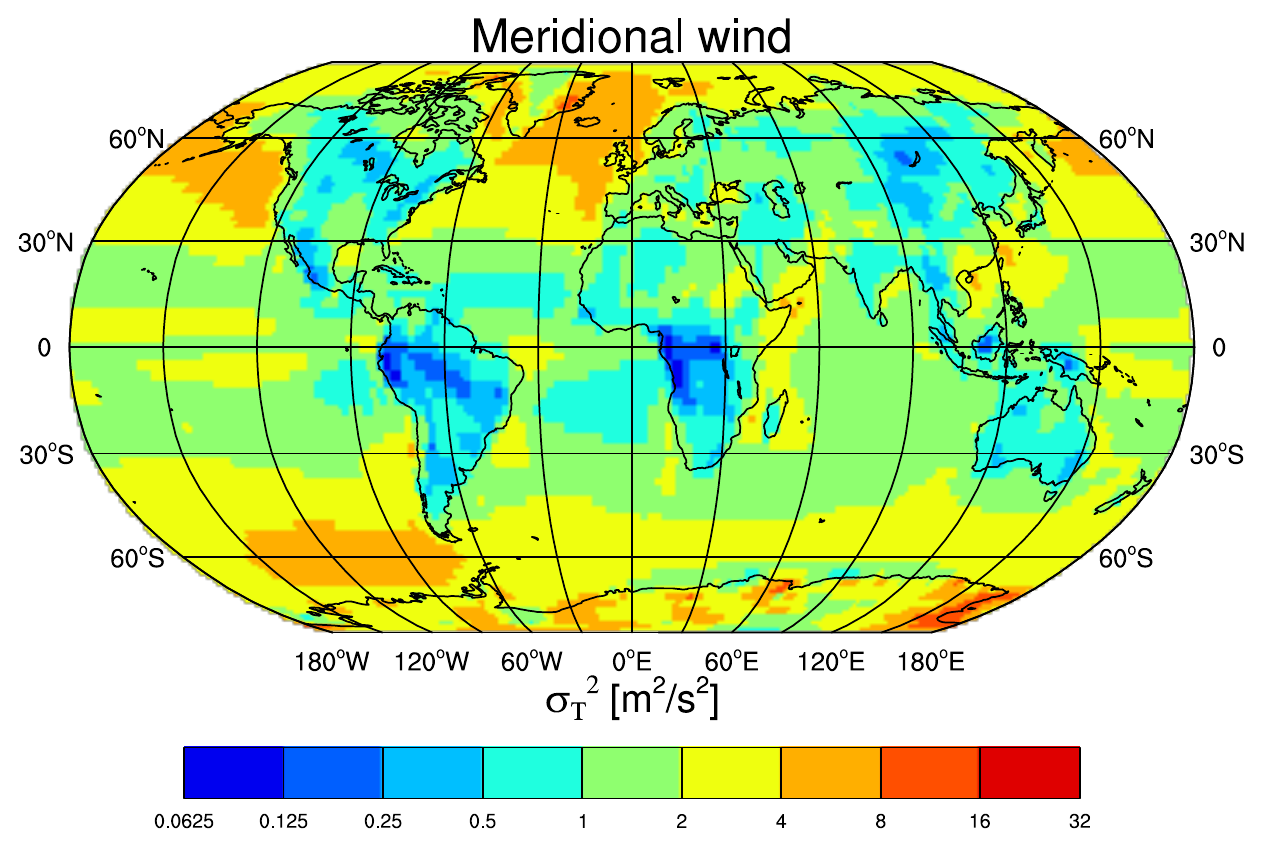}}
  \caption{Spatial distribution of the temporal average of the total variability, $\sigma_T^2$, (in log scale) of the annual mean $clm$ anomalies of surface meridional wind, $V^c$. The temporal average was calculated from the 30-year prediction period. High variability is observed mainly over the oceans.}\label{VAR_global_total_vas}
\end{figure}

\begin{figure}[ht]
 \centerline{\includegraphics[width=19pc]{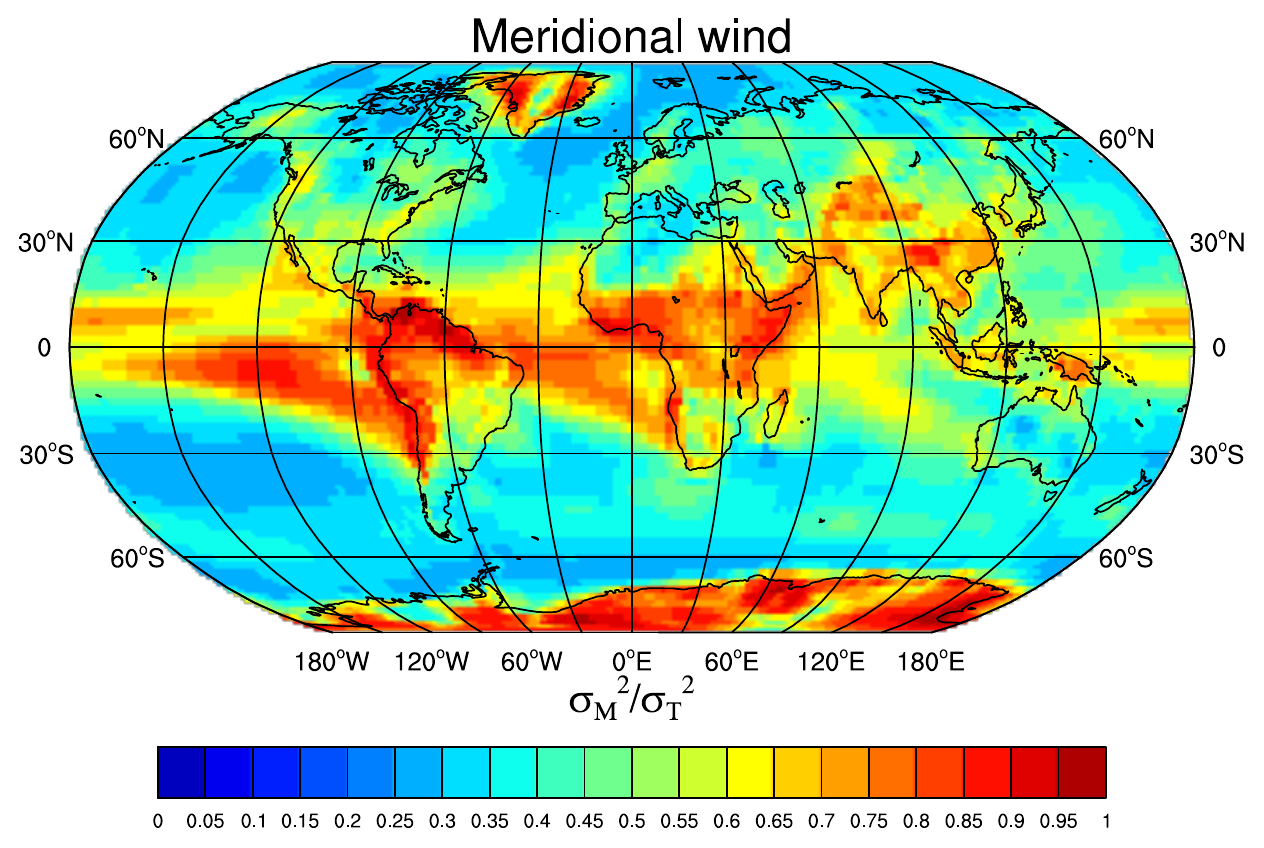}}
  \caption{Spatial distribution of the fraction of model variability from the total variability, $\sigma_M^2/\sigma_T^2$ , for $V^c$. Model variability is the main source of uncertainty over land, while the internal variability is larger over the oceans.}\label{VAR_global_model_frac_vas}
\end{figure}

\begin{figure*}[ht]
 \centerline{\includegraphics[width=39pc]{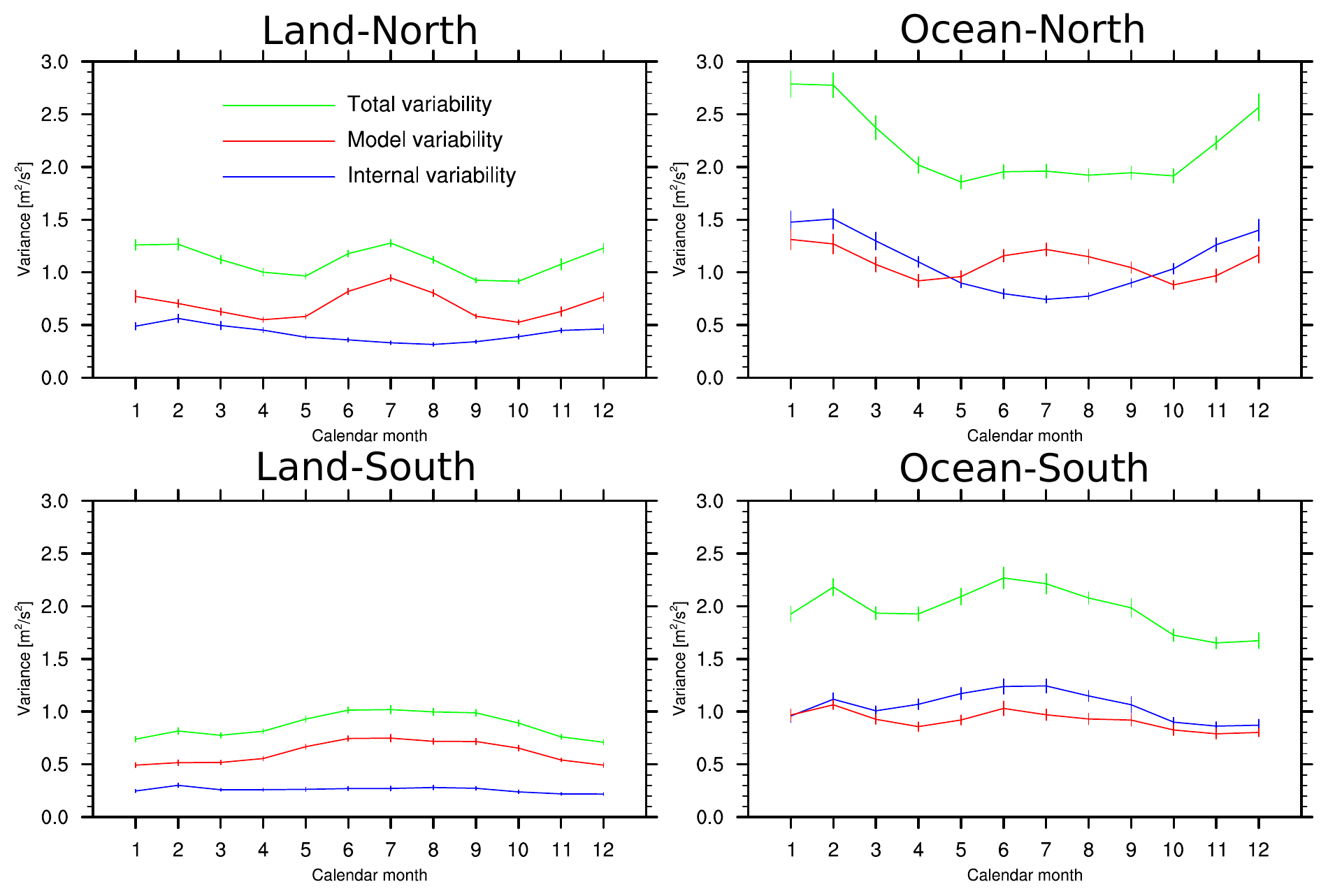}}
  \caption{Temporal average of the $V^c$ internal, model and total variabilities for four domains--over land in the northern hemisphere, over land in the southern hemisphere, over ocean in the northern hemisphere and over ocean in the southern hemisphere. The temporal averages were calculated from the 30-year prediction period for each calendar month. The error bars represent two standard deviations calculated from the 30-year time series of the global average variance for each month. The oceans in both hemispheres are the main sources of variability.}\label{VAR_vas_month}
\end{figure*}

This section presents the analysis of the surface meridional wind anomalies and their variabilities.
The analysis is similar to the analysis presented in the main text for the surface zonal wind.
Figure \ref{DATA_vas_year} presents the global and annual averages of the surface meridional wind anomalies. These anomalies are of the same order of magnitude as those found for the surface zonal wind.
Figure \ref{VAR_year_vas} presents the variabilities of the global and annual averages of $V$. The variabilities are a little smaller than those found for $U$. Figure \ref{VAR_month_vas} depicts the annual cycle of the surface meridional wind variabilities. Also for this variable, we found that the model variability is more influenced by the annual cycle. The maximal variability is found during the winter in the  northern hemisphere. However, for the bias-corrected anomalies, the peak is much lower due to the smaller contribution of the model variability. The spatial distribution of the total variability of $V^c$ and the fraction of the model variability are presented in Figures \ref{VAR_global_total_vas} and \ref{VAR_global_model_frac_vas}. The variability is larger over the oceans (similar to the surface zonal wind). Over land, the model variability is larger, while over the oceans, the internal variability dominates. Figure \ref{VAR_vas_month} presents the temporal average of the $V^c$ internal, model and total variabilities for four domains--over land in the northern hemisphere, over land in the southern hemisphere, over ocean in the northern hemisphere and over ocean in the southern hemisphere. The temporal averages were calculated from the 30-year prediction period for each calendar month. The error bars represent two standard deviations calculated from the 30-year time series of the global average variance for each month. The oceans in both hemispheres are the main sources of variability.

\section{Spatial distribution of the monthly mean variabilities for each calendar month}

\begin{figure*}[ht]
 \centerline{\includegraphics[width=33pc]{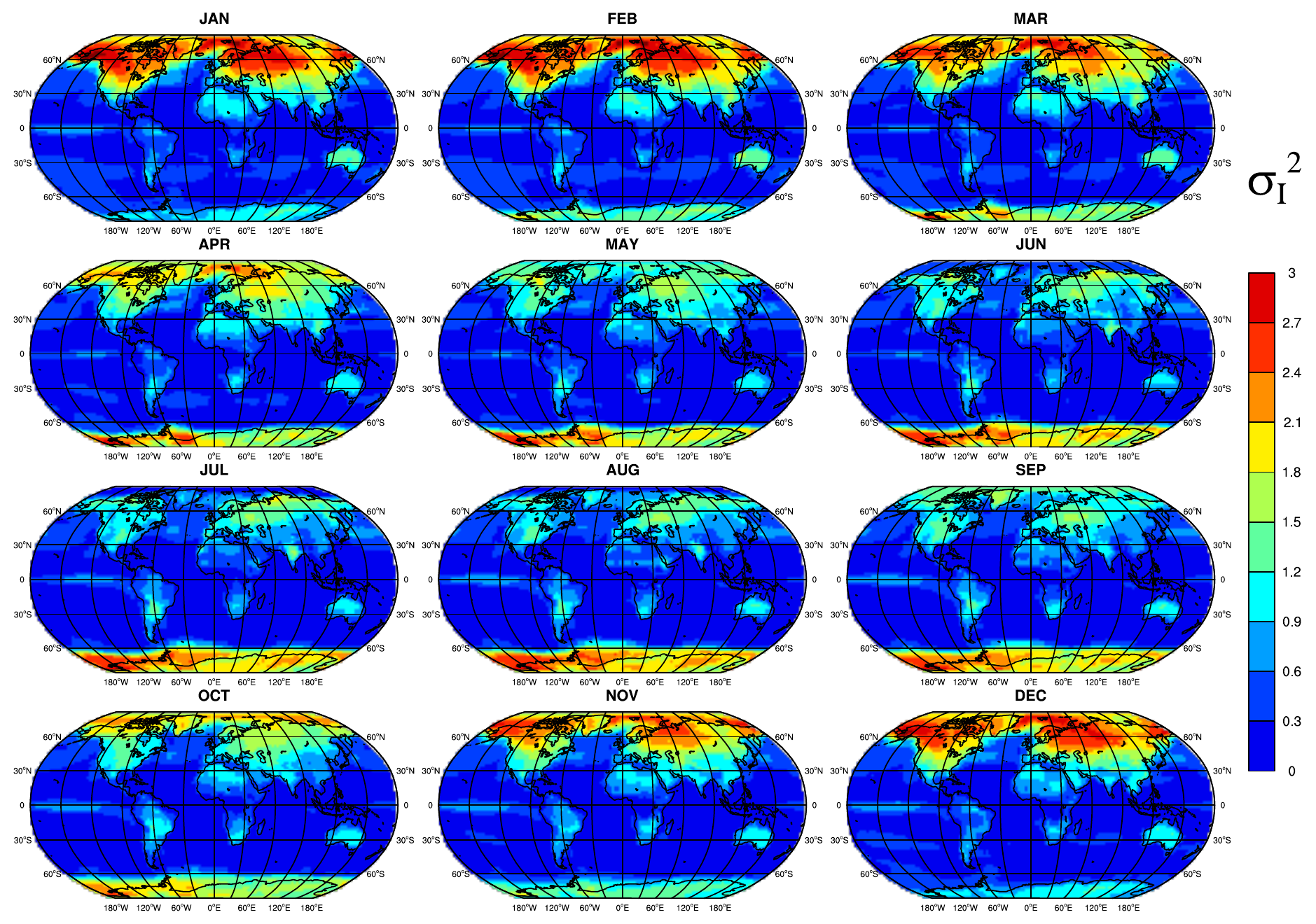}}
 \caption{Spatial distribution of $T^c$ internal variability for each calendar month. }\label{tas_nclm_VAR_global_internal_month}
\end{figure*}

\begin{figure*}[ht]
 \centerline{\includegraphics[width=33pc]{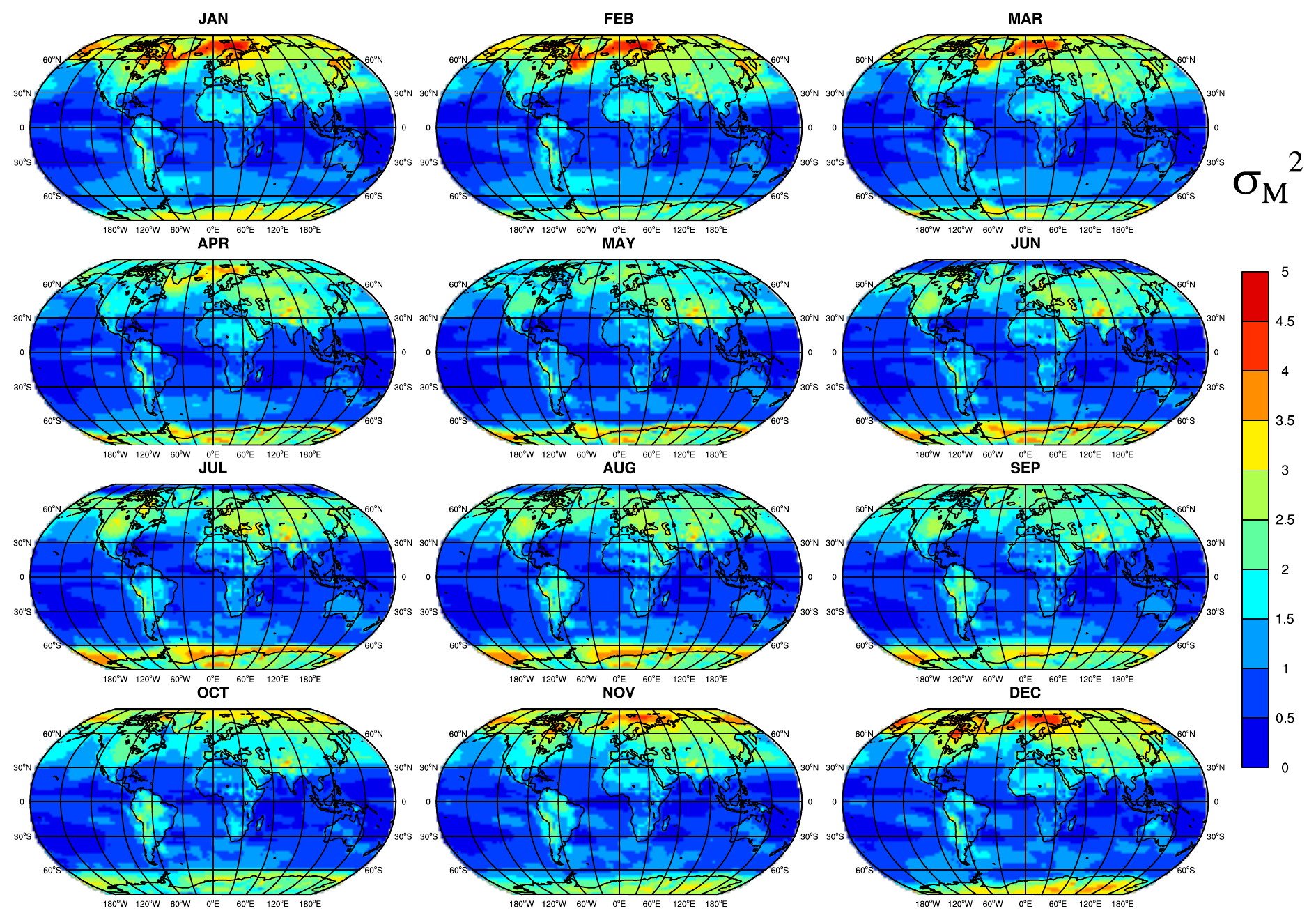}}
  \caption{Spatial distribution of $T^c$ model variability for each calendar month. }\label{tas_nclm_VAR_global_model_month}
\end{figure*}

\begin{figure*}[ht]
 \centerline{\includegraphics[width=33pc]{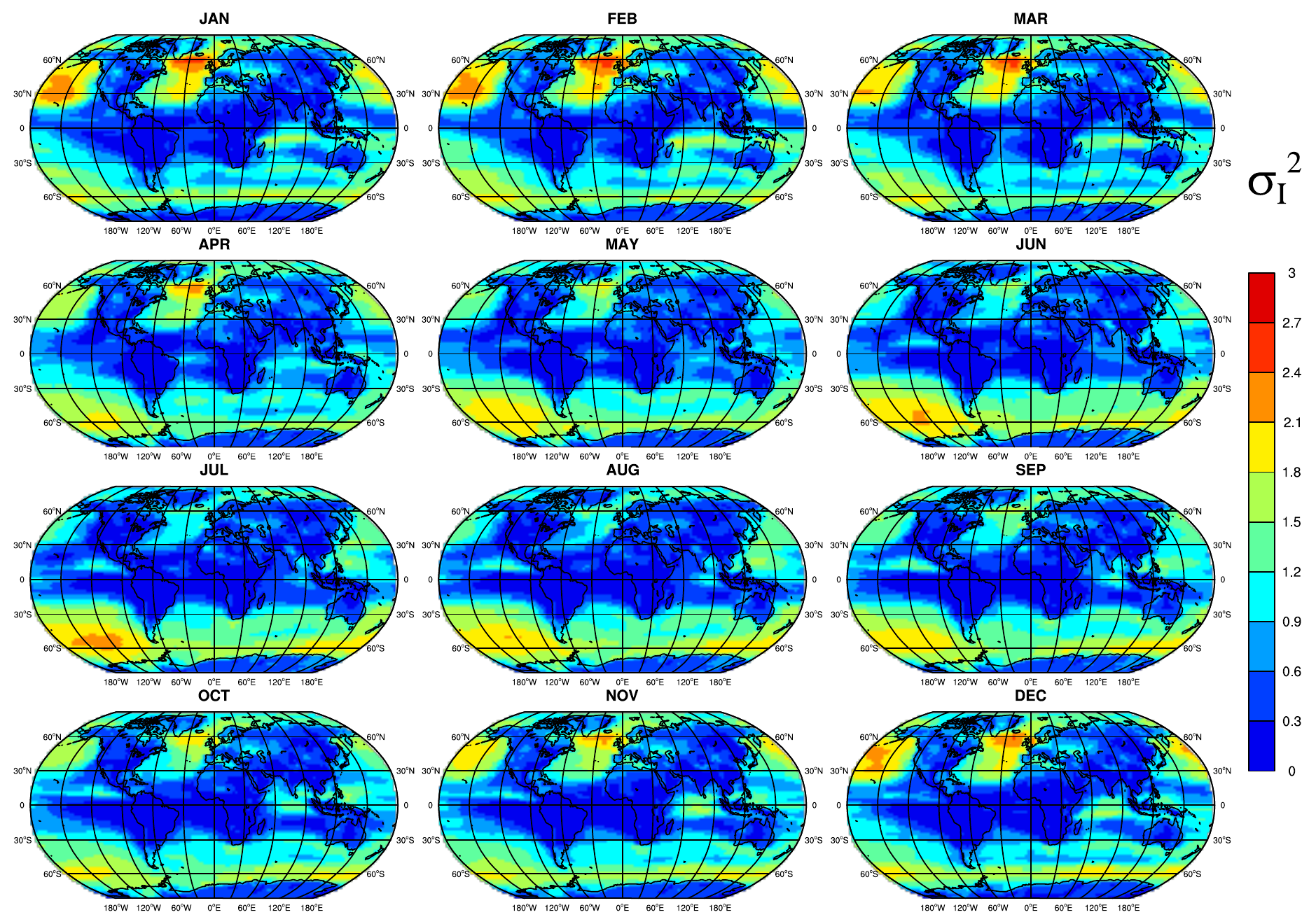}}
 \caption{Spatial distribution of $U^c$ internal variability for each calendar month. }\label{uas_nclm_VAR_global_internal_month}
\end{figure*}

\begin{figure*}[ht]
 \centerline{\includegraphics[width=33pc]{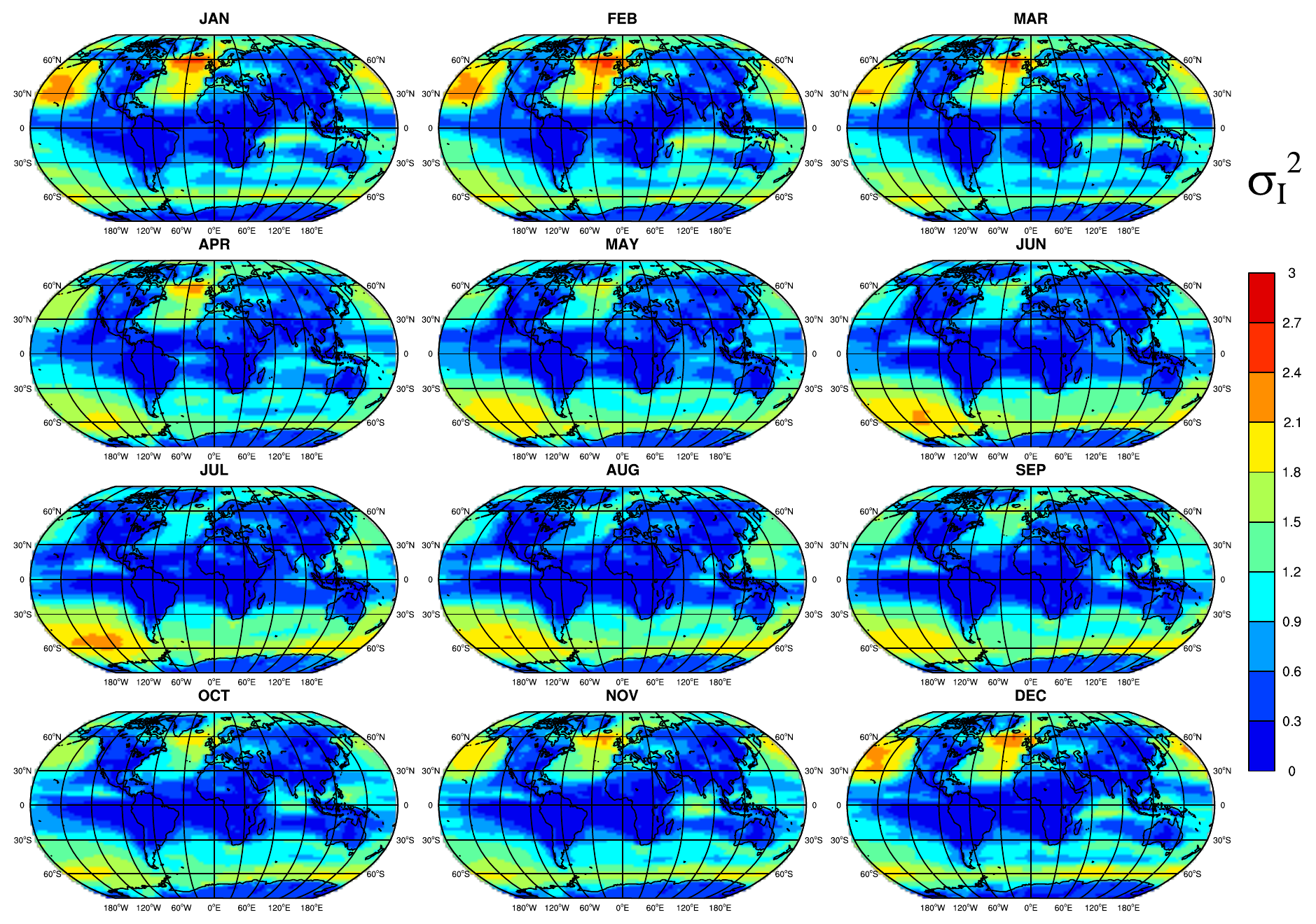}}
  \caption{Spatial distribution of $U^c$ model variability for each calendar month. }\label{uas_nclm_VAR_global_model_month}
\end{figure*}

\begin{figure*}[ht]
 \centerline{\includegraphics[width=33pc]{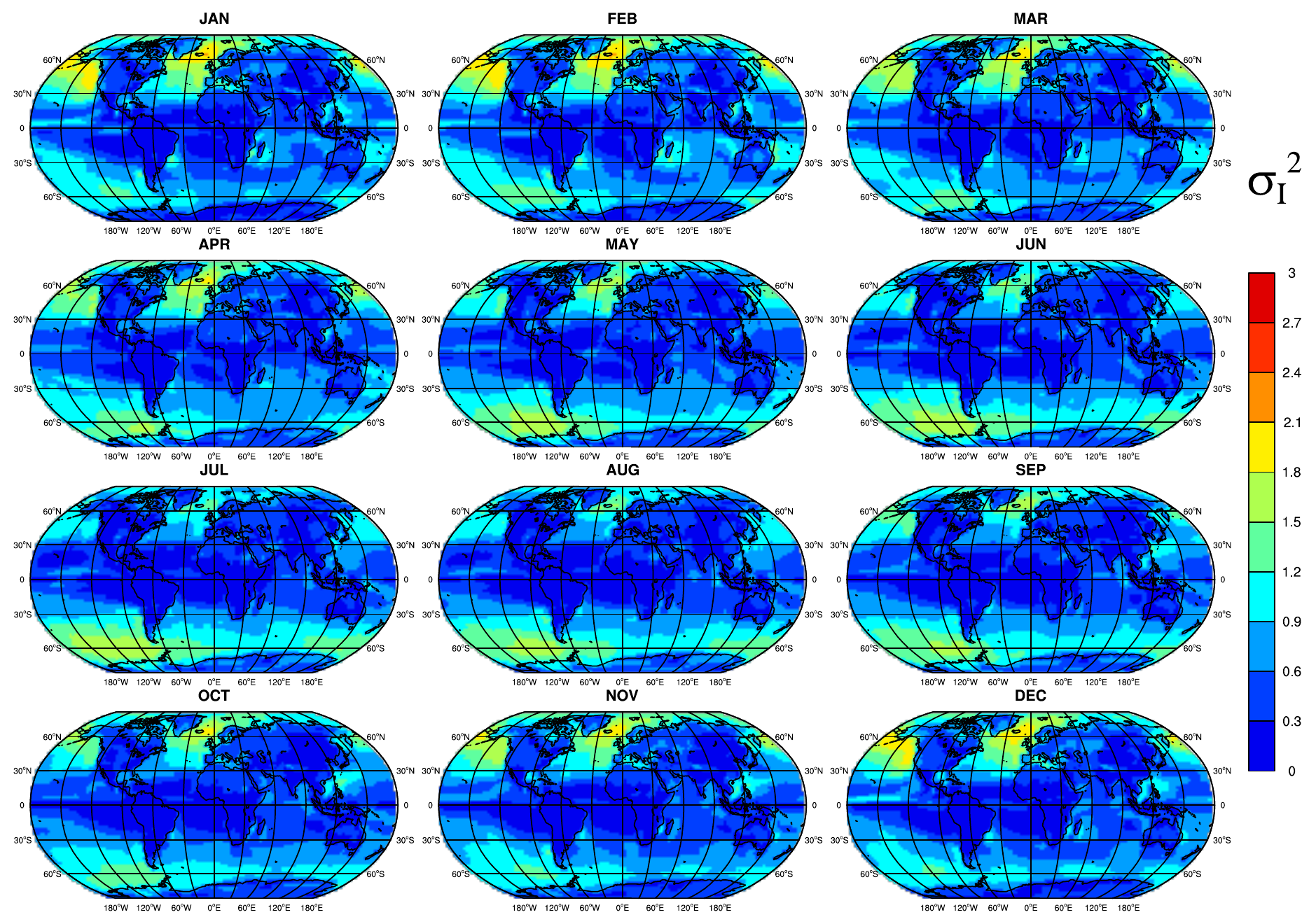}}
 \caption{Spatial distribution of $V^c$ internal variability for each calendar month. }\label{vas_nclm_VAR_global_internal_month}
\end{figure*}

\begin{figure*}[ht]
 \centerline{\includegraphics[width=33pc]{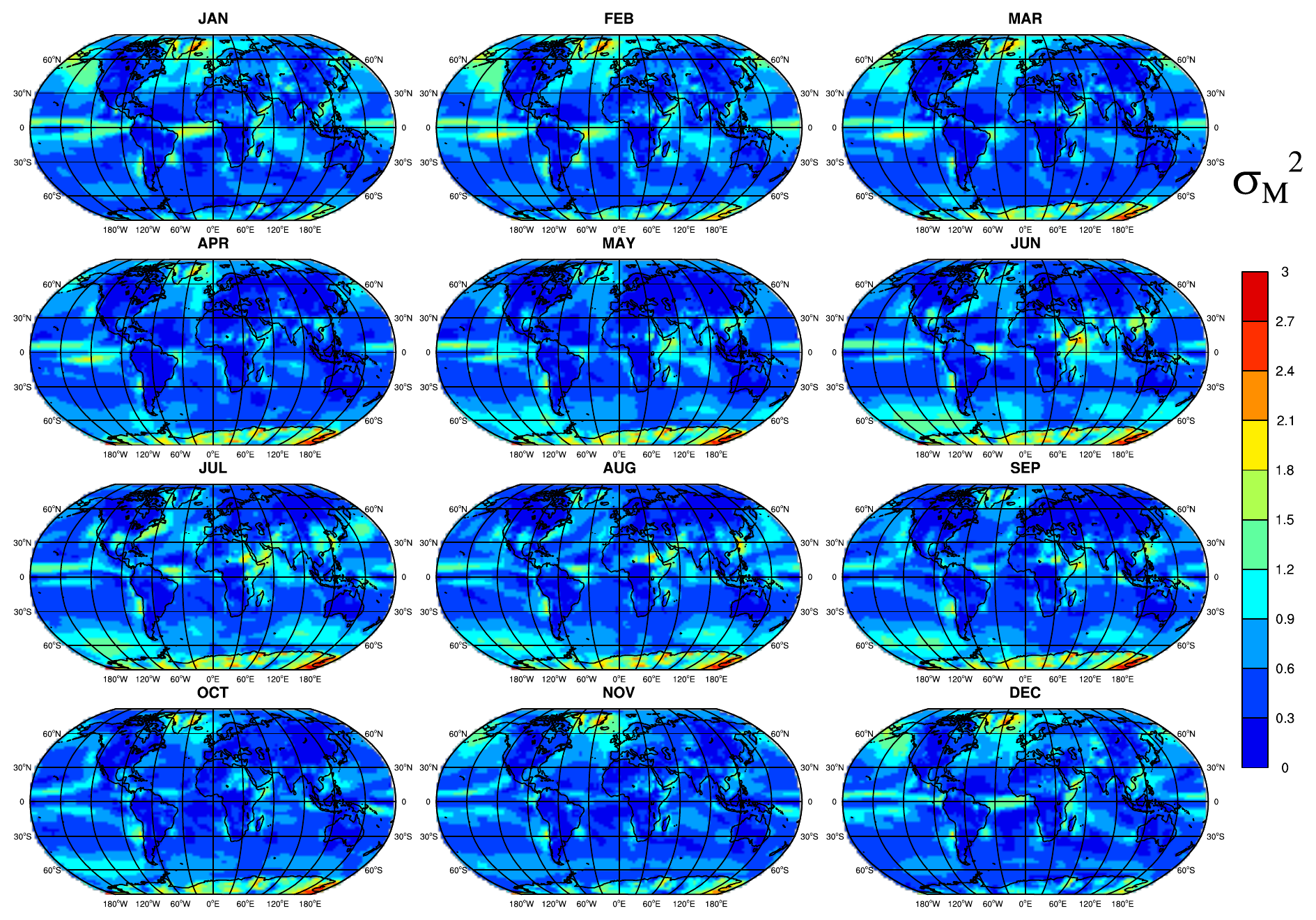}}
  \caption{Spatial distribution of $V^c$ model variability for each calendar month. }\label{vas_nclm_VAR_global_model_month}
\end{figure*}

In the main manuscript, the variabilities of the monthly means in four regions are presented. Here, we extend the analysis to show the full spatial distributions of the model and internal variabilities of the surface temperature, zonal wind and meridional wind $clm$ anomalies. The variabilities presented were derived from the temporal average, for each calendar month, of the 30-year prediction period.
Figures \ref{tas_nclm_VAR_global_internal_month} and \ref{tas_nclm_VAR_global_model_month} present the variabilities of the surface temperature anomaly.
Figures \ref{uas_nclm_VAR_global_internal_month} and \ref{uas_nclm_VAR_global_model_month} present the variabilities of the surface zonal wind anomaly, and Figures \ref{vas_nclm_VAR_global_internal_month} and \ref{vas_nclm_VAR_global_model_month} present the variabilities of the surface meridional wind anomaly.

\end{document}